\documentclass[onecolumn,11pt]{IEEEtran}
\IEEEoverridecommandlockouts

\usepackage{graphicx}
\usepackage{balance}
\usepackage[T1]{fontenc}
\usepackage{ifthen}
\usepackage[cmex10]{amsmath} 

\usepackage[lined,boxed,commentsnumbered,linesnumbered, ruled]{algorithm2e}
\usepackage[url,hyperrefblack,notheorems,IEEEtran]{research17}
\usepackage{comment}
\usepackage{booktabs} 
\usepackage{mathtools}
\usepackage{amsmath,amssymb,amsfonts,amsthm}
\usepackage{tikz} 
\tikzset{font={\fontsize{10pt}{12}\selectfont}}
\usetikzlibrary{shapes, patterns,decorations.text, decorations.pathreplacing,plotmarks}

\newtheorem{lemma}{\mylemmaname}
\newtheorem{theorem}{\mytheoremname}
\newtheorem{definition}{\mydefinitionname}

\newtheorem{corollary}{\mycorollaryname}
\newtheorem{example}{\myexamplename}
\newtheorem{remark}{\myremarkname}

\usepackage[capitalize]{cleveref}

\crefname{equation}{\unskip}{\unskip}
\crefname{claim}{Claim}{Claims} 
\usepackage{array}
\usepackage{multirow}
\newcolumntype{C}[1]{>{\centering\arraybackslash}p{#1}}
\allowdisplaybreaks

\makeatletter
\def\@xfootnote[#1]{%
	\protected@xdef\@thefnmark{#1}%
	\@footnotemark\@footnotetext}
\makeatother

\renewcommand{\vect}[1]{\vectg{#1}} 
\renewcommand{\vmat}[1]{\bm{\mat{#1}}} 
\newcommand{\code}[1]{\mathscr{#1}} 
\newcommand*{\Resize}[2][4]{\resizebox{#1}{!}{\ensuremath{#2}}} 
\makeatletter
\renewcommand*\env@matrix[1][*\c@MaxMatrixCols c]{%
	\hskip -\arraycolsep
	\let\@ifnextchar\new@ifnextchar
	\array{#1}}
\makeatother

\newcommand{\HP}[1]{\HH\left(#1\right)} 
 
\newcommand{\bigHP}[1]{\HH\bigl(#1\bigr)}

\newcommand{\HPcond}[2]{\HH\left(#1 \kern0.1em\middle|\kern0.1em #2\right)}
\newcommand{\eHPcond}[2]{\HH(#1 \kern0.1em|\kern0.1em #2)} 
\newcommand{\bigHPcond}[2]{\HH\bigl(#1 \kern-0.1em \bigm| \kern-0.1em#2\bigr)}
\newcommand{\BigHPcond}[2]{\HH\Bigl(#1 \kern-0.1em \Bigm| \kern-0.1em#2\Bigr)}
\newcommand{\MI}[2]{\II\left(#1 \kern0.1em{;}\kern0.1em #2\right)} 
\newcommand{\eMI}[2]{\II(#1 \kern0.1em{;}\kern0.1em #2)} 
\newcommand{\bigMI}[2]{\II\bigl(#1 \kern0.1em{;}\kern0.1em #2\bigr)}
\newcommand{\BigMI}[2]{\II\Bigl(#1 \kern0.1em{;}\kern0.1em #2\Bigr)}
\newcommand{\MIcond}[3]{\II\left(#1 \kern0.1em{;}\kern0.1em #2 \kern0.1em\middle|\kern0.1em #3\right)}
\newcommand{\eMIcond}[3]{\II(#1 \kern0.1em{;}\kern0.1em #2 \kern0.1em|\kern0.1em #3)} 
\newcommand{\bigMIcond}[3]{\II\bigl(#1 \kern0.1em{;}\kern0.1em #2 \kern-0.1em \bigm| \kern-0.1em#3\bigr)}
\newcommand{\BigMIcond}[3]{\II\Bigl(#1 \kern0.1em{;}\kern0.1em #2 \kern-0.1em \Bigm| \kern-0.1em#3\Bigr)}
\newcommand{\m}{\color{magenta}} 
\DeclareSymbolFont{matha}{OML}{txmi}{m}{it}
\DeclareMathSymbol{\varv}{\mathord}{matha}{118}
\usepackage{upgreek}
\usepackage{tablefootnote}
\usepackage{cases}

\begin{document}
\title{Pliable Private Information Retrieval} 

 \author{%
   \IEEEauthorblockN{Sarah A.~Obead and J{\"o}rg Kliewer\\
   \IEEEauthorblockA{Helen and John C.~Hartmann Department of Electrical and Computer Engineering \\ 
   					New Jersey Institute of Technology, Newark, New Jersey 07102, USA
   					\thanks{This work is supported in part by US NSF grants 1815322 and 2201824.}}
 }
}

\maketitle


\begin{abstract}
	
	We formulate a new variant of the private information retrieval (PIR) problem where the user is pliable, i.e., interested in \emph{any} message from a desired subset of the available dataset, denoted as pliable private information retrieval (PPIR). We consider a setup where a dataset consisting of $f$ messages is replicated in $n$ noncolluding databases and classified into $\Gamma$ classes. For this setup, the user wishes to retrieve \emph{any} $\lambda\geq 1$ messages from \emph{multiple} desired classes, i.e., $\eta\geq 1$, while revealing no information about the identity of the desired classes to the databases. We term this problem multi-message PPIR (M-PPIR) and introduce the single-message PPIR (PPIR) problem as an elementary special case of M-PPIR. %
		%
		%
		We first derive converse bounds on the M-PPIR rate, which is defined as the ratio of the desired amount of information and the total amount of downloaded information, followed by the corresponding achievable schemes.
		As a result, we show that the PPIR capacity, i.e., the maximum achievable PPIR rate, for  $n$ noncolluding databases  
		matches the capacity of PIR with $n$ databases and $\Gamma$ messages. Thus, enabling flexibility, i.e., pliability, where privacy is only guaranteed for classes, but not for messages as in classical PIR, allows to trade-off privacy versus download rate. A similar insight is shown to hold for the general case of M-PPIR. 
		
\end{abstract}


\section{Introduction}
\label{sec:introduction}

Today, a growing amount of traffic over the internet is generated by content-based applications. Content-based applications are applications that provide access to information (e.g., search engines, video libraries, and digital galleries) generated by individuals or businesses. Examples of well-known content-based applications include news-feed applications, social media, and content delivery networks.  
This prominent presence of content-type versus traditional message-type traffic in communication networks has recently caught the attention of the network information theory community.  For example, \cite{SongFragouli15_1} explored the benefits of designing network and channel codes tailored to content-type requests. 
The main distinction is that content-type traffic is able to deliver a message within a prescribed content type instead of specific messages.  

In this work, motivated by emerging content-based applications and inspired by content-type coding, 
we introduce the pliable private information retrieval (PPIR) problem as a new variant of the classical private information retrieval (PIR) problem.  
PIR was established originally in theoretical computer science by Chor \emph{et al.} \cite{ChorGoldreichKushilevitzSudan98_1} and has recently attracted much attention in the information and coding theory communities. As a result, many interesting variations of the PIR problem have surfaced (see e.g., \cite{ SunJafar17_1, ShahRashmiRamchandran14_1, BanawanUlukus18_1,ChanHoYamamoto15_1,Freij-HollantiGnilkeHollantiKarpuk17_1,Freij-Hollanti-etal19_1,KumarLinRosnesGraellAmat19_1,TajeddineGnilkeElRouayheb18_1,
	BanawanUlukus18_2,
	ChenWangJafar20_2,HeidarzadehGarciaKadheEl-RouayhebSprintson18_1, HeidarzadehSprintson22_1sub, KadheGarciaHeidarzadehElRouayhebSprintson20_1, LiGastpar20_2,ShariatpanahiSiavoshaniMaddah-Ali18_1, WeiBanawanUlukus19_2,
	BanawanUlukus19_1, SunJafar18_2, TajeddineGnilkeKarpukFreij-HollantiHollanti19_1, WangSunSkoglund19_1,
	SunJafar19_1,WangSkoglund19_2, WangSkoglund19_1 }).  
Such variants include additional privacy, storage and security constraints. For example, the fundamental limit of PIR from replicated distributed storage systems (DSSs), i.e., DSSs consisting of databases encoded with simple repetition codes, was presented in \cite{SunJafar17_1} while other coded storage scenarios were considered in \cite{ShahRashmiRamchandran14_1, BanawanUlukus18_1,ChanHoYamamoto15_1,Freij-HollantiGnilkeHollantiKarpuk17_1, Freij-Hollanti-etal19_1,KumarLinRosnesGraellAmat19_1,TajeddineGnilkeElRouayheb18_1}.
 In \cite{BanawanUlukus18_2}, multi-message PIR (M-PIR) has been proposed where the user can request more than one messages from replicated databases.
 Another interesting PIR variant where the user already knows a subset of the messages stored in the database, i.e., PIR with side-information, was studied in \cite{HeidarzadehGarciaKadheEl-RouayhebSprintson18_1, ShariatpanahiSiavoshaniMaddah-Ali18_1,LiGastpar20_2,ChenWangJafar20_2, KadheGarciaHeidarzadehElRouayhebSprintson20_1, HeidarzadehSprintson22_1sub,WeiBanawanUlukus19_2}. 
 In \cite{BanawanUlukus19_1,SunJafar18_2,TajeddineGnilkeKarpukFreij-HollantiHollanti19_1, WangSunSkoglund19_1} a bounded number of databases might be colluding, adversarial (byzantine), non-responsive, or eavesdropping. Finally, symmetric PIR where an additional privacy constraint is introduced to protect  database privacy, i.e., the user learns nothing about the dataset other than the desired message, was considered in \cite{SunJafar19_1,WangSkoglund19_2, WangSkoglund19_1}.

The PIR problem and its available variations traditionally aim to retrieve a \emph{specific} information message from a database without revealing the identity of the desired message to the database under a minimum communication cost.  This broad aim encompasses most of the work in the PIR literature.
However, in (single-message) PPIR,  we consider that the user is flexible with her demand. She wishes to retrieve \emph{any} message from a desired subgroup of the dataset, i.e., \emph{class},  without revealing the identity of the desired class to each database. This significantly distinguishes PPIR from classical PIR with two salient features: (i) The user does not know the identity of the messages in each class and only intends to keep the class index, but not the message index, private from the databases; (ii)  with each new instance of the protocol the answers are randomized among the messages in each class, if the same classes are queried by the user repetitively. Hence, existing PIR solutions and the corresponding capacity results, in general,  cannot  be immediately applied to the PPIR problem. We aim to fill this void in this paper.

One motivating example for PPIR is given by retrieving a news article of a desired topic without revealing the topic to the database.  
Another example would be to privately retrieve a movie from a desired genre without revealing the genre, i.e., the classification of the movie, to the content database  in order to avoid targeted recommendations or undesired profiling. In some cases, the user may be interested in retrieving more than one message from a number of desired classes, and that motivates the introduction of multi-message PPIR (M-PPIR) as a natural extension to the M-PIR and single-message PPIR problems.  Similarly to the PPIR motivating examples, the user might be interested in retrieving 
news articles from a number of popular desired topics or in retrieving a collection of movies from a set of desired genres. 
To illustrate the difference between PIR and PPIR, consider the following example.
\begin{example}\label{Ex:PPIR1}(Pliable Private Information Retrieval) Suppose that we have a single database consisting of $f=5$ equal-length messages denoted by $\vect{W}^{(1)},\dots, \vect{W}^{(f)}$, being classified into $\Gamma=2$ classes. Suppose that the messages with indices $\mathcal{M}_1=\{1,3\}$  are members of the first class $\gamma=1$ and the remaining messages, i.e.,  messages with indices $\mathcal{M}_2=\{2,4,5\}$ are members of the second class $\gamma=2$. Consider a user that is interested in retrieving \emph{any} message from class $\gamma=1$ while keeping the class index hidden from the database. If the user has access to the message membership in each class, i.e., the user knows $\mathcal{M}_1=\{1,3\}$ and  $\mathcal{M}_2=\{2,4,5\}$, there are two intuitive solutions.
	
	\begin{itemize}
		\item  One solution is to select one of the members of the desired class uniformly at random and attempt to privately retrieve that message using a PIR solution. For achieving information-theoretic privacy in the single-server case it is well-known that the user must  download the entire database to hide the identity of the desired  message \cite{ChorGoldreichKushilevitzSudan98_1}. As a result, the information retrieval rate, the ratio of the desired amount of information and the total amount of downloaded information, is given as $\const{R}=\frac{1}{f}=\frac{1}{5}$. 
		\item Alternatively, in PPIR the user selects $\Gamma$ messages uniformly at random, one from each class. Let the selected messages indices be denoted by $\theta_1$ and $\theta_2$, respectively, for each class. The user then queries the database for the two messages $\vect{W}^{(\theta_1)}$ and $\vect{W}^{(\theta_2)}$, resulting in probabilities ${\mathbb P}(\gamma=1|\theta_1,\theta_2)= {\mathbb P}(\gamma=2|\theta_1,\theta_2)=\frac{1}{\Gamma}$. In other words, perfect information-theoretic privacy is achieved as the desired message can be from any of the two classes. As a result the information retrieval rate is given as  $\const{R}=\frac{1}{\Gamma}=\frac{1}{2}$. This matches the PIR rate for the case where we have only $f=2$ messages stored in the database, indicating an apparent trade-off between the  reduction of message privacy and the download rate.
	\end{itemize}
\end{example}

It can be seen from \Cref{Ex:PPIR1} that the PPIR rate reduces to the PIR rate if there is only one message in each class, i.e., $\Gamma=f$. Accordingly, the PPIR problem is also a strict generalization of the PIR problem.
Moreover, we are able to achieve a significant gain in the information retrieval rate with the PPIR solution if $f\gg \Gamma$. Note that in  PPIR  we assume the the user is oblivious about the message memberships of each class. In contrast, the traditional PIR solution in \Cref{Ex:PPIR1} is not valid if the user does not know the identity of the messages that belong to the desired class.

To the best of our knowledge the problem of pliable private information retrieval has not been studied before in the open literature. 
However, there has been some related work on other PIR variations that explore trading off perfect message privacy with a privacy leakage to decrease the download rate. 
The following are some representative examples: 
\cite{SamyTandonLazos19_1} initiated the study of \emph{leaky} PIR for an arbitrary number of messages and two replicated databases and derived upper and lower bounds on the download rate for some bounded  $\epsilon>0$ information leakage on the message identity. 
Further, in weakly-private information retrieval  \cite{LinKumarRosnesGraellAmatYaakobi21_1,LinKumarRosnesGraellAmatYaakobi22_1,GuoZhouTian20_1,QianZhouTianLiu2022_app}, the perfect privacy requirement on the identity of the desired message is relaxed by allowing bounded  average leakage between the queries and the corresponding requested message index.  The leakage is  measured by using different information leakage measures including mutual information and maximum leakage  \cite{Smith09_1,BartheKopf11_1,IssaWagnerKamath20_1}. In particular, \cite{LinKumarRosnesGraellAmatYaakobi21_1,LinKumarRosnesGraellAmatYaakobi22_1} studied the trade-offs between different parameters of PIR, such as download rate, upload cost, and access complexity while relaxing the privacy requirement. 

Another related line of research, inspired by content-type coding \cite{SongFragouli15_1}, is given by \emph{pliable index coding (PICOD)} \cite{BrahamaFragouli15_1} as a variant of the classical Index coding (IC) problem \cite{BirkKol98,BirkKol06}. IC is a well-known network information theory problem that shares an intimate connection to the problem of PIR with side information. In IC the aim is to minimize the broadcast rate for communicating of messages  noiselessly 
to $n$  receivers, where each 
receiver has a different subset of messages as side information. PICOD is a variant of the IC problem where the receivers, having a set of messages as side information, are interested in \emph{any} other message they do not have. This is in contrast to classical IC, where the receivers are interested in \emph{specific} messages. 
Following the introduction of PICOD, converse bounds on the PICOD broadcast rate were derived in  \cite{LiuTuninetti20_2}. Moreover, variations of the PICOD problem are considered in  \cite{LiuTuninetti19_1,LiuTuninetti19_2, LiuTuninetti20_1}. 
Specifically, in private PICOD \cite{LiuTuninetti19_2}, the privacy is defined by the inability of each user to decode more than one message.
In decentralized PICOD  \cite{LiuTuninetti19_1}, the system model departs from the assumption of a  central transmitter with knowledge of all $f$ messages. Here, the $n$ users share messages among themselves which can only depend on their local set of side information messages. This work has been recently extended to secure decentralized PICOD in \cite{LiuTuninetti20_1} where security is defined such that users are not allowed to gain  information about any message outside their side information set except for one message.  Finally, a number of constructions for PICOD are proposed in \cite{LiuTuninetti21_1, JiangShi18_1,Song18_1 , SasiRajan19_1, OngVellambiKliewer19_1, OngVellambi2022_app,  KrishnanMathewKalyanasundaram21_1}.

\subsection{Main Contributions}

In this paper, we introduce the multi-message PPIR (M-PPIR) problem where we solely focus on downloading from $n$ noncolluding replicated  databases.
Our contributions are outlined as follows:

\begin{itemize}   
	\item First, we fully characterize the PPIR capacity where the user is interested in downloading one message from one desired class. These findings are later extended to the general M-PPIR case, where the user intends to download multiple messages from an arbitrary subset of classes. 
	\item Towards this end,  we prove a novel converse bound on the M-PPIR rate for an arbitrary number of messages $f$, classes $\Gamma,$ and databases $n$ and we construct a capacity-achieving PPIR scheme.
	The significance of our derived converse bounds is that in contrast to PIR  they indicate an independence between the maximum achievable rate and the total number of files $f$. 
	When there is only one message in each class, i.e., $\Gamma=f$, the M-PPIR problem reduces to the M-PIR problem and our converse bounds match the M-PIR bounds.
	
	\item Finally, by leveraging our achievable scheme for  PPIR  and the M-PIR schemes of  \cite{BanawanUlukus18_2}, we present two achievable M-PPIR constructions. The first scheme applies to the case when number of desired classes by the user is at least half the total number of classes $\eta\geq \frac{\Gamma}{2}$ and the second when $\eta\leq \frac{\Gamma}{2}$. The achievable rates of the proposed schemes match the converse bounds when $\eta\geq \frac{\Gamma}{2}$ and when $\frac{\Gamma}{\eta}$ is an integer number. Thus, we settle the M-PPIR capacity from replicated databases for  these two cases.
\end{itemize}

 The reminder of the paper is organized as follows. In \Cref{sec:preliminaries}, we outline the notation and formally define the M-PPIR problem. In \Cref{sec:PPIR}, we derive the converse bound for single-message PPIR as special case of M-PPIR and present an achievable scheme that matches the converse.
 In \cref{sec:MPPIR}, we consider the general case of M-PPIR and derive upper and lower bounds on its capacity along with an example.  \Cref{sec:MPPIR-conclusion} offers the conclusion.

\section{Preliminaries}\label{sec:preliminaries}

\subsection{Notation}
\label{sec:notation}

We denote by $\Naturals$ the set of all positive integers and for some $a,b \in\Naturals$,
$[a]\eqdef\{1,2,\ldots,a\}$ and $[a:b]\eqdef\{a,a+1,\ldots,b\}$ for $a\leq b$ . 
A random variable is denoted by a capital Roman letter, e.g., $X$, while its realization is denoted by the corresponding small
Roman letter, e.g., $x$. Vectors are boldfaced, e.g., $\vect{X}$ denotes a random vector and $\vect{x}$ denotes a
deterministic vector, respectively. 
In addition, sets are denoted by calligraphic upper case letters, e.g., $\set{X}$. 
For a given index set $\set{S}$, we also write $\vect{X}^\set{S}$ and $Y_\set{S}$ to represent $\bigl\{\vect{X}^{(v)}\colon v\in\set{S}\bigr\}$ and $\bigl\{Y_j\colon j\in\set{S}\bigr\}$, respectively. 
Furthermore, some constants and functions are depicted by Greek letters or a special font, e.g., $\const{X}$. 
$\trans{(\cdot)}$ denotes the transpose operator, 
$\HP{X}$ represents the entropy of $X$, and $\MI{X}{Y}$ the mutual
information between $X$ and $Y$. 
 $\mathbb P[A]$ is the
probability that the event $A$ occurs.

\subsection{System Model}
\label{sec:system-model}

We consider a dataset that consists of a number of $f$ independent messages $\vect{W}^{(1)},\ldots,\vect{W}^{(f)}$. Each message $\vect{W}^{(m)}=\bigl(W_{1}^{(m)},\dots,W_{\const{L}}^{(m)}\bigr)$, $m\in [f]$, is a random length-$\const{L}$ vector for some $\const{L} \in\Naturals$, with independent and identically distributed symbols that are chosen at random from the field $\Field_p$. The messages are classified into $\Gamma$ classes for $\Gamma\leq f$\footnote[$\dagger$]{Note that we assume that every message is classified into one class only and no class is empty, i.e., $\Gamma \ngtr f$. }, $\Gamma \in \mathbb{N}$, and replicated in a distributed storage system (DSS) consisting of $n$ noncolluding databases.  Without loss of generality, we assume that the symbols of each message are selected uniformly over the field  $\Field_p$. Thus,
\begin{IEEEeqnarray}{rCl}
	\bigHP{\vect{W}^{(m)}}& = &\const{L},\,\forall \,m\in[f], \label{eq:Entropy}
	\\
	\bigHP{\vect{W}^{(1)},\dots,\vect{W}^{(f)}}& = & { f\const{L}\quad (\textnormal{in } p\textnormal{-ary units}).} \label{eq:JointEntropy}
\end{IEEEeqnarray}

Let $\mathcal{M}_{\gamma}$ be the set of \emph{message indices} belonging to the class indexed with $\gamma\in[\Gamma]$ where $M_{\gamma}=|\mathcal{M}_{\gamma}|$  is the size of this set.
Note that here we assume that every message is classified into one class only i.e., $\forall \gamma', \gamma\in [\Gamma]$ and $\gamma'\neq \gamma,$ $\mathcal{M}_\gamma \cap \mathcal{M}_{\gamma'} =\phi$  and $\sum_{\gamma=1}^{\Gamma} M_\gamma=f.$ Moreover, we assume that there are at least two classes, i.e., $1\leq M_\gamma \leq f-1$. Finally, for simplicity of presentation and without loss of generality, we assume that messages are ordered in an ascending order based on their class membership with  $ \mathcal{M}_\gamma = [(1+\sum_{i=1}^{\gamma-1}M_{i}):(\sum_{i=1}^{\gamma}M_i)]$ for all $\gamma\in [\Gamma]$, 
i.e., 
\begin{IEEEeqnarray*}{rCl}
	\{\vect{W}^{(1)},\dots, \vect{W}^{(M_1)}\} &\in& \vect{W}^{\mathcal{M}_1},\\
	\{\vect{W}^{(M_1+1)}, \dots, \vect{W}^{(M_1+M_2)} \} &\in& \vect{W}^{\mathcal{M}_2}, \\
	&\vdots&\\
	\{\vect{W}^{(1+\sum_{i=1}^{\Gamma-1}M_{i}) }, \dots, \vect{W}^{(f) } \} &\in& \vect{W}^{\mathcal{M}_\Gamma}.
\end{IEEEeqnarray*}%

\begin{figure}[t!]
	\centering
	\includegraphics[scale=0.23]{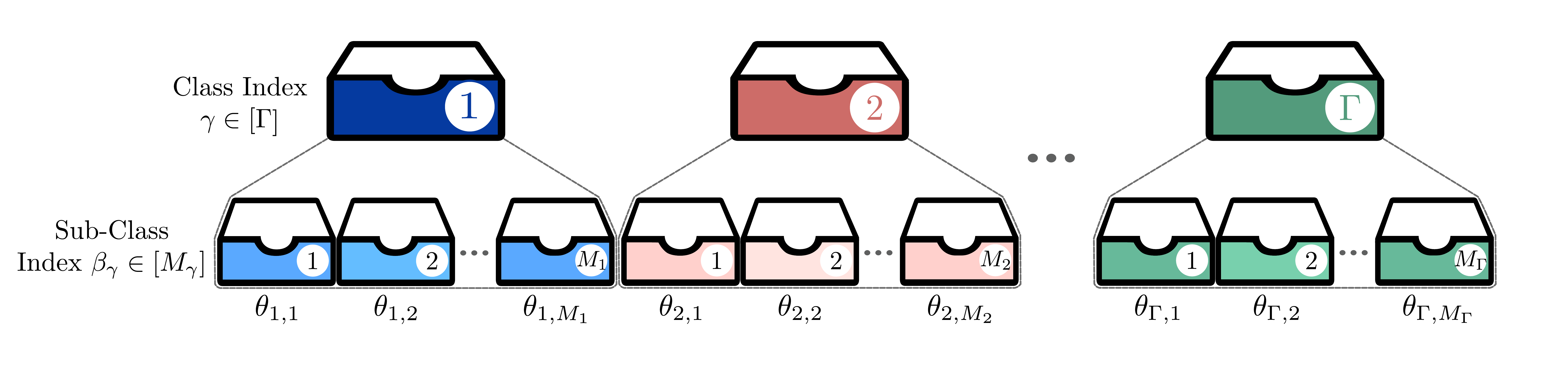}
	\caption{Index-mapping of $f$ messages classified into $\Gamma$ classes using class and sub-class indices, i.e., $\theta_{\gamma,\beta_{\gamma}}\in {\mathcal M}_{\gamma}\subset [f], $ $\forall \gamma\in [\Gamma].$}
	\label{fig:PPIR-IndexMapping}
\end{figure}

To represent the message index-mapping that results from  classifying the $f$ messages into $\Gamma$ classes, let, for $\gamma\in[\Gamma]$, $\theta_{\gamma,\beta_{\gamma}} $ be the index of a message that belongs to class $\gamma$ where $\beta_{\gamma} \in [M_{\gamma}]$ is a sub-class index and  $\theta_{\gamma,\beta_{\gamma}} \in \mathcal{M}_{\gamma}$. 
	 Here, the sub-class index $\beta_{\gamma}$ represents the membership of a message \emph{within} the class $\gamma$ as shown in \Cref{fig:PPIR-IndexMapping}.
Hence,  $\forall \gamma\in [\Gamma]$ and  $\forall \beta_{\gamma} \in  [M_{\gamma}]$, we have the index-mapping 
	\begin{IEEEeqnarray}{rCl} \label{eq:indexMapping}
		\theta_{\gamma,\beta_{\gamma}}\triangleq \beta_{\gamma}+ \sum_{l=1}^{\gamma-1} M_l.
	\end{IEEEeqnarray}%

\begin{example}\label{ex:indexMappingEx1}
 Assume that the messages with indices $\{9, 10, 11\}\subset [f]$ are members of the second class, i.e., $\mathcal{M}_2=\{9,10,11\}$ and $M_2=3$. Then, $\vect{W}^{(\theta_{2,1})}=\vect{W}^{(9)}$, $\vect{W}^{(\theta_{2,2})}=\vect{W}^{(10)}$, and $\vect{W}^{(\theta_{2,3})}=\vect{W}^{(11)}$. 
\end{example}

\subsection{Problem Statement}\label{sec:ProblemStatment}

\begin{figure}[t!]
	\centering
	\includegraphics[scale=0.21]{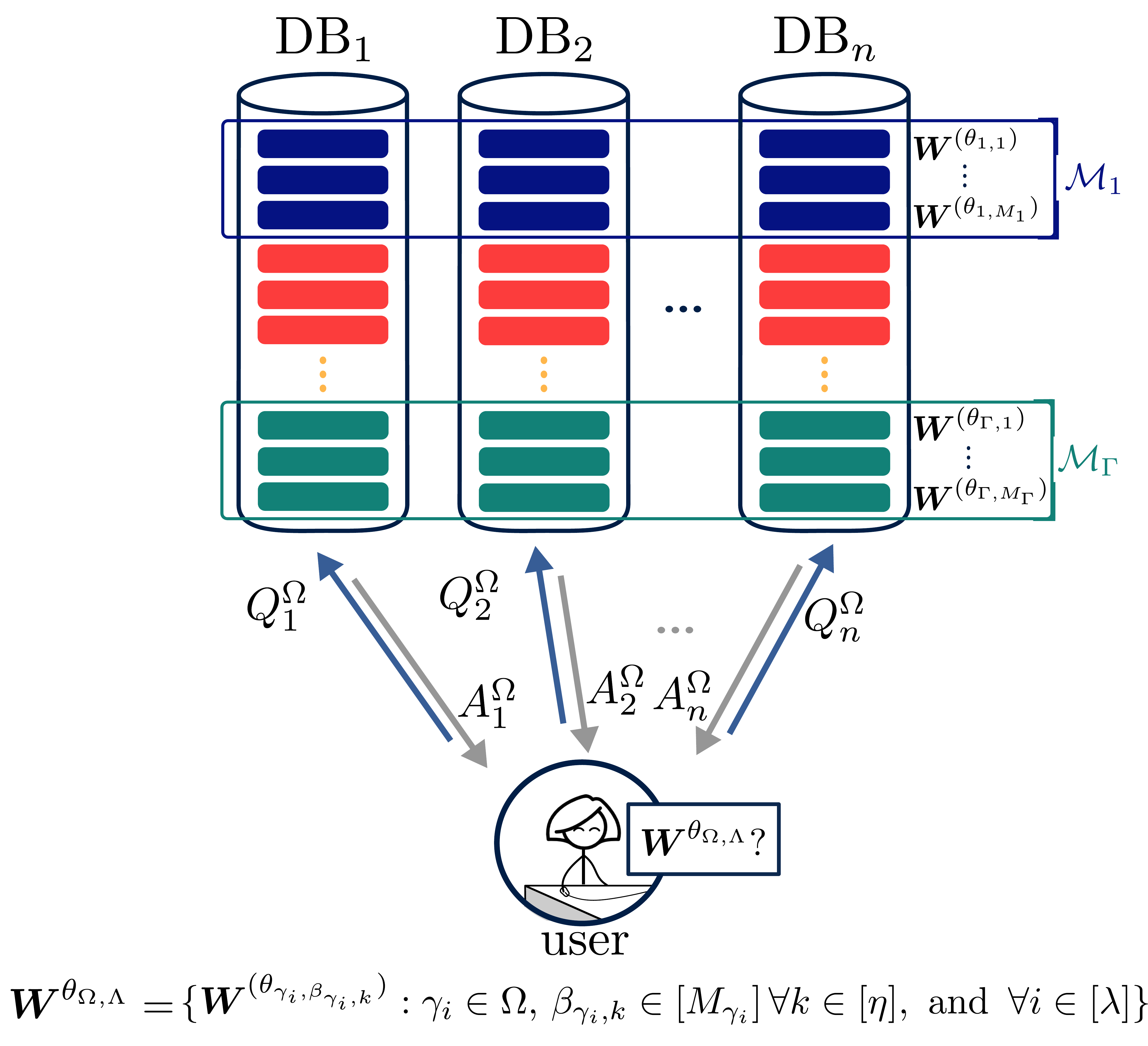}
	\caption{System model for M-PPIR from an $n$ replicated noncolluding databases storing $f$ messages classified into $\Gamma$ classes. The user intends to download $\lambda$ messages each out of $\eta$ desired classes.}
	\label{fig:PPIR-SysModel}
\end{figure}

In the multi-message PPIR (M-PPIR) problem, the user wishes to retrieve a total of \emph{any} $\mu$ messages from a subset of $\eta$ \emph{desired} classes indexed by the index set $\Omega \subseteq [\Gamma]$ where $|\Omega|=\eta$. The desired number of messages $\mu$ is distributed among the desired classes as $\mu = \sum_{i=1}^{\eta}\lambda_{\gamma_i}$ where $\lambda_{\gamma_i}$ is the number of \emph{desired} messages from the desired class $\gamma_i \in \Omega$. %
For the scope of this work  and for tractability we restrict ourselves to a fixed number of requested messages from each desired class, i.e., $\lambda_{\gamma_i}=\lambda  \; \forall \gamma_i \in \Omega$ and $\mu=\lambda\eta$. Moreover, we impose the mild assumption that the user only has prior knowledge of the least common multiple (LCM) of the sizes of the $\Gamma$ classes $\delta \eqdef \text{LCM}(M_1,\dots,M_\Gamma)$. In other words, the user \emph{does not} know the \emph{size} of each class or the total number of files stored at the database.
Accordingly, the user wishes to privately retrieve \emph{any} $\lambda$ messages out of $M_{\gamma_i}$ messages within a desired \emph{class}  $\gamma_i\in \Omega$, $\forall i\in [\eta]$,  which are denoted by $\{\vect{W}^{(\theta_{\gamma_1,\beta_{\gamma_1,1}})}, \vect{W}^{(\theta_{\gamma_1, \beta_{\gamma_1,2}})},\dots, \vect{W}^{(\theta_{\gamma_1, \beta_{\gamma_1,\lambda}})}, \dots, \vect{W}^{(\theta_{\gamma_\eta,\beta_{\gamma_\eta,\lambda}})}\}$,
i.e.,  
\begin{IEEEeqnarray*}{rCl}
	\{\vect{W}^{(\theta_{\gamma_i,\beta_{\gamma_i,k}})}: {\gamma_i} \in \Omega,\, \beta_{\gamma_i,k}\in [M_{\gamma_i}] \quad \forall k \in [\lambda], \text{ and }\,\forall  i\in [\eta]\}.
\end{IEEEeqnarray*}

\begin{figure}[t!]
	\centering
	\includegraphics[scale=0.17]{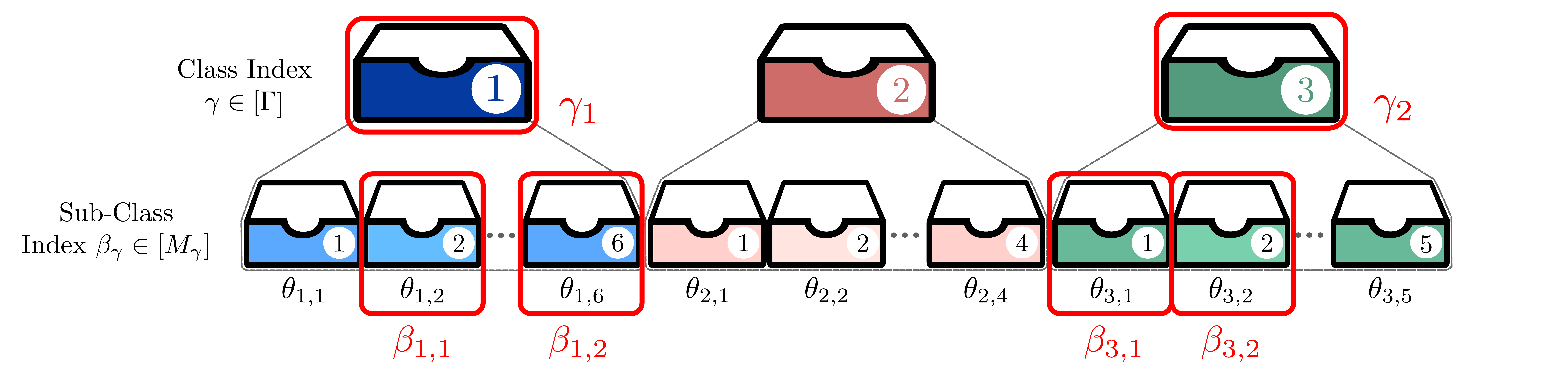}
	\caption{Index mapping for M-PPIR problem of \Cref{ex:indexMappingEx2}. The user selects $\Omega=\{1,3\}$, i.e., $\gamma_1=1$ and $\gamma_2=3$ and wants to retrieve any two messages from each class. Highlighted in red, are two arbitrary sub-class indices from each desired class.}
	\label{fig:PPIR-IndexMappingEX}
\end{figure}

\begin{example}\label{ex:indexMappingEx2}
	Consider a dataset consisting of $f=15$ messages classified into $\Gamma=3$ classes with sizes $\{6,4,5\}$, respectively.  Suppose a user that wishes to retrieve any $\lambda=2$ messages from the set of classes $\Omega=\{1,3\}$. The indices of the \emph{two} arbitrary selected messages from each class are shown in \Cref{fig:PPIR-IndexMappingEX}. The sub-class index of the \emph{first} message from the \emph{first} class, i.e., $i=1$, $k=1$, and $\gamma_1=1$, is given by $\beta_{1,1}=2$. From the index-mapping of \eqref{eq:indexMapping}, we have $\theta_{1,\beta_{1,1}}= \beta_{1,1}= 2$ and similarly, $\theta_{1,\beta_{1,2}}= \beta_{1,2}=M_1=6$. Next, the sub-class index of the \emph{first} message from the \emph{second} class, i.e., $i=2$, $k=1$ and $\gamma_2=3$, is given by $\beta_{3,1}=1$. From the index-mapping  \eqref{eq:indexMapping}, we have $\theta_{3,\beta_{3,1}}= 1+ \sum_{l=1}^{2} M_l= 11$ and similarly for  $\theta_{3,\beta_{3,2}}= 2+ \sum_{l=1}^{2} M_l=12$.
\end{example}

The user privately selects a subset of $\eta$ class indices $\Omega =\{\gamma_1, \gamma_2, \dots, \gamma_\eta\} \subseteq [\Gamma]$ and wishes to retrieve \emph{any} $\lambda$ messages from each of the desired classes, while
keeping the identities of the requested classes in $\Omega$ private from each database. In order to retrieve the desired messages  $\{\vect{W}^{(\theta_{\gamma_1,\beta_{1,1}})},\dots, \vect{W}^{(\theta_{\gamma_1, \beta_{1,\lambda}})},\dots, \vect{W}^{(\theta_{\gamma_\eta, \beta_{\eta,\lambda}})}\}$,  the user sends a random query $Q^{\Omega}_j$ to the database $j\in [n]$. The query is generated by the user without any prior knowledge of the realizations of the stored messages. In other words, 
\begin{IEEEeqnarray}{rCl}
	\MI{\vect{W}^{(1)},\ldots,\vect{W}^{(f)}}{Q^{\Omega}_1, \dots,Q^{\Omega}_n }=0. \label{eq:IndepQM}
\end{IEEEeqnarray}
In response to the received query, the $j$-th database sends the answer $A^{\Omega}_j$ back to the user, where  $A^{\Omega}_j$ is a
deterministic function of $Q^{\Omega}_j$ and the data stored in the database. Thus, 
\begin{IEEEeqnarray}{rCl}
\bigHPcond{A^{\Omega}_j}{Q^{\Omega}_j,\vect{W}^{[f]}}=0,\; \forall\,j\in [n]. \label{eq:DeterministicAnswers}
\end{IEEEeqnarray}

Note that, here we assume that there exists \emph{at least} $\lambda$ messages in each class, i.e., $M_\gamma\geq \lambda, \, \forall \gamma\in [\Gamma]$. Let $\mathcal{V}$ and $\mathcal{T}$ be two arbitrary subsets of $\mathcal{M}_\gamma$ such that $\mathcal{V}\subseteq \mathcal{T}\subseteq {\mathcal M}_\gamma$ and $|\mathcal{V}|=\lambda$. It follows from the definition of the M-PPIR problem that
\begin{IEEEeqnarray}{rCl}\label{eq:AnswersFromPartialDS}
	\bigHPcond{A^{\Omega}_j}{Q^{\Omega}_j,\vect{W}^{\mathcal V}}=  \bigHPcond{A^{\Omega}_j}{Q^{\Omega}_j,\vect{W}^{\mathcal T}}.
\end{IEEEeqnarray}

This is unlike the classical PIR setup where the answer string is generated given all of the messages in the dataset. Hence, from the chain rule of entropy we have  $\bigHPcond{A^{\Omega}_j}{Q^{\Omega}_j,\vect{W}^{\mathcal V}}\geq   \bigHPcond{A^{\Omega}_j}{Q^{\Omega}_j,\vect{W}^{\mathcal T}}$  for the classical M-PIR.
In other words, in M-PPIR, the answer from the database $j \in [n]$ is generated as a deterministic function given a sufficient amount of information, i.e., at least \emph{any} $\lambda$ messages from a class for any class $\gamma \in[\Gamma]$. Similarly, let $v'\in\cset{{\mathcal M}}_\gamma \eqdef [f]\setminus {\mathcal M}_\gamma $ and $\set{V'} \subseteq \cset{{\mathcal M}}_\gamma$. Then it follows from \eqref{eq:AnswersFromPartialDS} that
\begin{IEEEeqnarray}{rCl}
	\bigHPcond{A^{\Omega}_j}{Q^{\Omega}_j,\vect{W}^{\mathcal V}\vect{W}^{(v')}}=  \bigHPcond{A^{\Omega}_j}{Q^{\Omega}_j,\vect{W}^{\mathcal T}\vect{W}^{(v')}}
\end{IEEEeqnarray}
and 
\begin{IEEEeqnarray}{rCl}
	\bigHPcond{A^{\Omega}_j}{Q^{\Omega}_j,\vect{W}^{\mathcal V}\vect{W}^{\mathcal V'}}=  \bigHPcond{A^{\Omega}_j}{Q^{\Omega}_j,\vect{W}^{\mathcal T}\vect{W}^{\mathcal V'}}.
\end{IEEEeqnarray}

To satisfy the class  privacy requirement, the query-answer function must be identically distributed for all possible subset of class indices
$\Omega \subseteq [\Gamma]$ from the perspective of each database. In other words, the scheme's query and answer string must be independent from the desired class index set, i.e., 
\begin{IEEEeqnarray}{rCl}
	\MI{\Omega}{Q^{\Omega}_j, A^{\Omega}_j}=  0, \, \forall j\in [n].\label{eq:MPPIR-privacyALT}
\end{IEEEeqnarray}
 Moreover, the user must be able to reliably decode, given the received databases answers, any $\lambda$ messages from the desired classes i.e., $\{\vect{W}^{(\theta_{\gamma_1,\beta_{\gamma_1,1}})},\dots,  \vect{W}^{(\theta_{\gamma_1,\beta_{\gamma_1,\lambda}})},\dots, \vect{W}^{(\theta_{\gamma_\eta,\beta_{\gamma_\eta,\lambda}})} \} $ for $\gamma_i\in \Omega$. Accordingly, the M-PPIR protocol from replicated DSS is defined as follows.

Consider a DSS with $n$ noncolluding replicated databases storing $f$ messages classified into $\Gamma$ classes. The user wishes to retrieve any $\lambda$ messages from each class in the desired class index set $\Omega \subseteq[\Gamma]$, from the queries $Q^{\Omega}_j$ and answers $A^{\Omega}_j$, $\forall j\in[n]$. Let $\mathfrak{S}$ be the set of all unique subsets of $[\Gamma]$ of size $\eta$, and $\mathcal{M}_{\gamma_i}$ be the index set of the messages classified into the class $\gamma_i\in \Omega$, then for an M-PPIR
protocol, the following conditions must be satisfied $\forall\,\Omega,\Omega' \in  \mathfrak{S}$, $\Omega \neq \Omega'$, and $j\in [n]$:
\begin{IEEEeqnarray}{rCl}    
	&&\textnormal{[Privacy]} \quad  \quad  \;\; \; (Q^{\Omega}_j,A^{\Omega}_j,\vect{W}^{[f]})  \sim (Q^{\Omega'}_j,A^{\Omega'}_j,\vect{W}^{[f]})\footnote{ The privacy constraint can be alternatively expressed as eq.~\eqref{eq:MPPIR-privacyALT}.}, \label{eq:MPPIR-privacy}
	\\
	&&\textnormal{[Correctness]}  \quad   \bigHPcond{\vect{W}^{(\theta_{\gamma_1,\beta_{\gamma_1,1}})},\dots, \vect{W}^{(\theta_{\gamma_1,\beta_{\gamma_1,\lambda}})},\dots \vect{W}^{(\theta_{\gamma_\eta,\beta_{\gamma_\eta,\lambda}})}}{A^{\Omega}_{[n]},Q^{\Omega}_{[n]}}=0.  
 	 \label{eq:MPPIR-correctness}
\end{IEEEeqnarray}

We summarize the important variables of the M-PPIR problem in \Cref{tab:notation}. 
	
	\begin{table}[h!]  
		\centering
		\caption{Important variables} 
		\label{tab:notation}
		\Resize[0.94\columnwidth]{
			\begin{IEEEeqnarraybox}[
				\IEEEeqnarraystrutmode
				\IEEEeqnarraystrutsizeadd{3pt}{2pt}]{v/c/v/s/V/c/v/s/v}
				\IEEEeqnarrayrulerow\\
				& \text{Notation} && Description												  && \text{Notation}  && 	Description\\ 	\hline\hline
				& n				&& 	 total number of databases (integer)							&& \Omega 	&& 	set of desired classes 	  \\*\hline
				& f 			 &&   total number of messages (integer)							&&\eta	&&number of desired classes (integer)	\\*\hline
				& \Gamma 		  && total number of classes (integer) 						&&\lambda   && 	number of desired messages from each desired class (integer) \\*\hline
				& \const{L} &&   number of symbols in each message (integer)				&&  M_\gamma	&& 	size of class $\gamma\in [\Gamma]$  (integer)\\*\hline
				& \mathfrak{S} &&   the set of all unique subsets of $[\Gamma]$ of size $\eta$		&& 	&& 	
				\\*\IEEEeqnarrayrulerow
			\end{IEEEeqnarraybox}
		}
	\end{table}

\subsection{Performance Metric}
To measure the efficiency of an M-PPIR protocol, we consider the required number of downloaded symbols for retrieving the $\const{L}$ symbols of the $\mu=\lambda\eta$ desired messages.

\begin{definition}[M-PPIR rate and capacity for replicated DSSs]
	\label{def:def_PCrate}
	The rate of an M-PPIR  protocol, denoted by $\const{R}$, is defined as the ratio of the desired information size, $\lambda\eta$ messages each consisting of 
	$\const{L}$ symbols, to the total required download cost $\const{D}$, i.e.,
	\begin{IEEEeqnarray*}{c}
		\const{R}\eqdef\frac{\eta\lambda\const{L}}{\const{D}} = \frac{\eta\lambda\const{L}}{\sum_{j=1}^{n} 	\bigHP{A^{\Omega}_j}}.
	\end{IEEEeqnarray*}
	The M-PPIR  \emph{capacity}, denoted by $\const{C}_\textnormal{M-PPIR}$, is the maximum achievable M-PPIR rate over all possible M-PPIR protocols.
\end{definition}

\subsection{Special Cases}
In this subsection, we introduce {two} 
special cases of the general M-PPIR problem presented in \cref{sec:system-model} emerging from choosing different values of $\lambda$ and $\eta$. 
We use these special cases, namely PPIR 
and multi-class PPIR, as building-blocks for the general M-PPIR problem. 	
As this work introduces 
the PPIR problem, we find it useful to see how these special cases relate to and extend classical PIR problems.


\subsubsection{Single-Message PPIR (in short denoted as PPIR $(\lambda=1,\eta=1)$)} \label{sec:PS:MC-PPIR} %
 Here, the user is interested in a \emph{single} message from a \emph{single} desired class\footnote{For notation simplicity, we drop the desired class subscript when it is understood from the context, e.g., if there is only one desired class $\eta=1$.}.   
In PPIR, the user privately selects a class index $\gamma\in[\Gamma]$ and wishes to privately retrieve \emph{any one} message out of the $M_{\gamma}$ \emph{candidate} messages of the desired class, i.e., 
 $\vect{W}^{(\theta_{\gamma,\beta_{\gamma,1}})}: \theta_{\gamma,\beta_{\gamma,1}} \in \mathcal{M}_{\gamma_1}, \, \gamma \in [\Gamma],$
while keeping the desired class index $\gamma$ private from each database $j\in [n]$. 
Note that when the number of classes is equal to the number of messages, i.e., there is only one message in each class and $\Gamma=f$, the PPIR problem reduces to the classical PIR problem \cite{SunJafar17_1}. %




\subsubsection{Multi-Class PPIR ($\lambda=1,\eta\geq 1$)}  \label{sec:PS:MC-MPPIR}	
Here, the user is interested in a \emph{single} message from \emph{multiple} desired classes. In this case, the user privately selects a subset of class indices $\Omega\subseteq[\Gamma]$ of size $\eta$ and wishes to retrieve \emph{any} one message from each of the $\eta$ desired classes $\gamma_i\in \Omega$, i.e.,  
 $\{\vect{W}^{(\theta_{\gamma_{1},\beta_{\gamma_1,1}})},\dots, \vect{W}^{(\theta_{\gamma_{\eta},\beta_{\gamma_\eta,1}})}:  \theta_{\gamma_i,\beta_{\gamma_i, 1}} \in \mathcal{M}_{\gamma_i},\, {\gamma_i} \in \Omega, \, \forall  i\in [\eta]\},$
without revealing the identity of the desired class index set  $\Omega$ to each database $j\in [n]$.
Note that when the number of classes is equal to the number of messages, i.e., there is only one message in each class and $\Gamma=f$, the multi-class PPIR problem reduces to the  multi-message PIR (MPIR) problem \cite{BanawanUlukus18_2}.

\section{Pliable Private Information Retrieval}
\label{sec:PPIR}

In this section, we discuss the PPIR problem as a special case of the M-PPIR problem with $\lambda=1, \eta=1$.  
The significance of presenting this special case lies within the direct connection to the well known classical PIR problem in \cite{SunJafar17_1}, thus, providing an intuitive introduction to the general M-PPIR problem. In the following, we derive the capacity of PPIR, %
 which indicates a significant (possible) reduction in download rate compared to the capacity of classical PIR. 
In the PPIR problem we assume that the user is \emph{oblivious} to the structure of the database, i.e., has no knowledge of the messages membership in each class and construct achievable schemes accordingly. 
To this end, we characterize the capacity of PPIR from replication-based DSSs and present a capacity-achieving scheme. Note that the novelty of our result is mostly in the converse proof, whereas the achievable scheme is based on a modified version of the scheme in \cite{SunJafar17_1}.  
We state our main result for PPIR over replicated DSS with \cref{thm:MS-PPIR} as follows.
\begin{theorem}	\label{thm:MS-PPIR}
	Consider a DSS with $n$ noncolluding replicated databases storing $f$ messages classified into $\Gamma$ classes.   The maximum achievable PPIR rate over all possible PPIR protocols, i.e., the PPIR capacity $\const{C}_{\textnormal{PPIR}}$, is given by
	\begin{IEEEeqnarray*}{rCl} 
		\const{C}_{\textnormal{PPIR}}
		&& = \inv{\left(1+ \frac{1}{n}+ \frac{1}{n^2}+\dots + \frac{1}{n^{\Gamma-1}} \right)}
		=  \left(1-\frac{1}{n}\right)  \inv{\left(1-\frac{1}{n^{\Gamma}}\right)}.
	\end{IEEEeqnarray*}
\end{theorem}

 	\begin{remark}\label{rem:SingleServerPPIR}
 		For $n=1$, the PPIR capacity  $\const{C}_{\textnormal{PPIR}}$ is given by $\const{C}_{\textnormal{PPIR}}= \frac{ 1}{\Gamma}.$
 		This can be shown by an induction argument. First, for $M_\gamma=1, \, \forall \gamma\in[\Gamma]$, i.e., each class contains only one message, we have $\Gamma=f$. Accordingly, in order to maintain the privacy of the desired class identity $\gamma \in [\Gamma]$, we must maintain the privacy of the retrieved message identity $\theta_{\gamma,1} \in [f]$. As a result, the capacity of single server PPIR matches the capacity of the single server PIR problem, i.e., $	\const{C}_{\textnormal{PPIR}}=\frac{1}{f}=\frac{1}{\Gamma}$ \cite{ChorGoldreichKushilevitzSudan98_1}.
 		Next, for $M_\gamma>1, \, \forall \gamma\in[\Gamma]$, in order to maintain the privacy of the desired class identity, we must download \emph{at least} one message from each class. Hence, the probability that any one of the classes is the desired class is uniformly distributed, thus achieving perfect information theoretic privacy. Since there is more than one message in each class, and the user requests \emph{any} message from her desired class, the identity of the selected message is not relevant.  Accordingly, by randomly selecting one message from each class as an answer to the user it follows that the best information retrieval rate, i.e., PPIR capacity, must be bounded by $\frac{1}{\Gamma}$, i.e., $\const{C}_{\textnormal{PPIR}}=\frac{1}{\Gamma}$.
 \end{remark} 

Before we start the converse proof of \Cref{thm:MS-PPIR}, we present a number of useful lemmas and simplifying assumptions. Without loss of generality, assume that: 
\begin{itemize}
	\item From the queries and answers of each database $j\in[n]$ we can successfully decode the first $\lambda$ messages in each desired class $\gamma_i\in \Omega$ for any $\Omega\in  \mathfrak{S}$,  where $\mathfrak{S}$ is the set of all unique subsets of $[\Gamma]$ of size $\eta$. As a result, $\beta_{\gamma_i,k}=k$ for all $k\in [\lambda],  i\in [\eta]$, and we can write the message index $\theta_{\gamma_i,\beta_{\gamma_i,k}}$ as $\theta_{\gamma_i,k}$ for all $k\in [\lambda]$. Let $\theta_{\gamma_i,k}$ denote the index of the $k$-th message in class $\gamma_i \in [\Gamma]$.
	Then, for example, from the answers of the desired classes indexed with set $\Omega=[\eta]=\{1,2,\dots,\eta\}$ we can successfully decode $\{\vect{W}^
	{(\theta_{1,1})}, \dots, \vect{W}^{(\theta_{1,\lambda})},\vect{W}^{(\theta_{2,1})}, \dots, \vect{W}^
	{(\theta_{\eta,\lambda})}\}$. For simplicity, with some abuse of notation, we let $\vect{W}^{\theta_{[\eta],[\lambda]} } \triangleq \{\vect{W}^
{(\theta_{1,1})}, \dots, \vect{W}^{(\theta_{1,\lambda})},\vect{W}^{(\theta_{2,1})}, \dots, \vect{W}^
{(\theta_{\eta,\lambda})}\} $. As a result from \eqref{eq:MPPIR-correctness} we have 
	$ \bigHPcond{\vect{W}^{\theta_{[\eta],[\lambda]}}}{A^{[\eta]}_{[n]},Q^{[\eta]}_{[n]}}=0$. 
	\item  Let $\vect{W}^{[f]\setminus\theta_{[\eta],[\lambda]}}$ be the complement subset of files for the set $\vect{W}^{\theta_{[\eta],[\lambda]}}$, where
	$$\theta_{[\eta],[\lambda]} \triangleq \{\theta_{1,1}, \theta_{1,2},\dots, \theta_{1,\lambda},\theta_{2,1}, \dots, \theta_{\eta,1},\dots, \theta_{\eta,\lambda}\},$$ i.e., 
	$\vect{W}^{[f]\setminus{\theta_{[\eta],[\lambda]}}} \triangleq $ $\vect{W}^{[\theta_{1,\lambda+1}:\theta_{2,1}-1]}\cup \vect{W}^{[\theta_{2,\lambda+1}:\theta_{3,1}-1]}\cup \dots \cup \vect{W}^{[\theta_{\eta-1,\lambda+1}:\theta_{\eta,1}-1]}\cup\vect{W}^{[\theta_{\eta,\lambda+1}:f]}$.
\end{itemize}

\begin{lemma} \label{lem:lem11} 
	$	\MIcond{ \vect{W}^{[f]\setminus\theta_{[\eta],[\lambda]}}}{Q^{[\eta]}_{[n]} A^{[\eta]}_{[n]}}{\vect{W}^{\theta_{[\eta],[\lambda]}}} 
	\leq \eta\lambda \const{L}(\frac{1}{\const{R}}-1).
	$
\end{lemma}
The proof of \cref{lem:lem11} is given in Appendix~\ref{app:lem1}.

\begin{lemma} \label{lem:lem22} 
	Let $\Omega_1, \Omega_2 \in  \mathfrak{S}$, such that $\Omega_1 \cap \Omega_2=\phi$,  without loss of generality, assume that $\Omega_1=[\eta]$ and $\Omega_2=[\eta+1:2\eta]$. Then
	\begin{IEEEeqnarray}{lCr}
		\MIcond{   \vect{W}^{[f]\setminus\theta_{[\eta],[\lambda]}} }{Q^{\Omega_1}_{[n]} A^{\Omega_1}_{[n]}}{    \vect{W}^{\theta_{[\eta],[\lambda]} }  }  %
		&\geq  \frac{\eta\lambda\const{L}}{n}   +\frac{1}{n}  \MIcond{\vect{W}^{ [f]\setminus \theta_{[2\eta],[\lambda]}  }}{Q^{\Omega_2}_{[n]} A^{\Omega_2}_{[n]}}{\vect{W}^{ \theta_{[2\eta],[\lambda]}} }. \label{eq:LowerBoundOnMI_1}
	\end{IEEEeqnarray}
\end{lemma}
The proof of \cref{lem:lem22} is given in Appendix~\ref{app:lem2}.

\subsection{Converse proof of \cref{thm:MS-PPIR}}  \label{sec:ConverseProof_MS-PPIR}
%
We now proceed to the converse proof of \Cref{thm:MS-PPIR}. For $\gamma\in[\Gamma]$, let $$\vect{W}^{\theta_{[\gamma],1}} \triangleq \{ \vect{W}^{(\theta_{1,1})}, \vect{W}^{(\theta_{2,1})},\dots, \vect{W}^{(\theta_{\gamma,1})} \}.$$ 
\begin{proof}		
	 
   From \Cref{lem:lem11} we have for $\lambda=1$ and $\eta=1$ 
   	\begin{IEEEeqnarray}{lCr}
   	\MIcond{\vect{W}^{[\theta_{1,2}:f]}}{Q^{(1)}_{[n]} A^{(1)}_{[n]}}{\vect{W}^{(\theta_{1,1})}} 	\leq  \const{L}\left(\frac{1}{\const{R}}-1\right).  \label{eq:lem1}
   	\end{IEEEeqnarray}
   Next, from \Cref{lem:lem22} we have for $\lambda=1$, $\eta=1$, and $\gamma \in{[2:\Gamma]}$ 
   	\begin{IEEEeqnarray}{lCr}
   			\MIcond{\vect{W}^{ [f]\setminus\theta_{[\gamma-1],1}  }  }{Q^{(\gamma-1)}_{[n]} A^{(\gamma-1)}_{[n]}}{\vect{W}^{ \theta_{[\gamma-1],1} }  }  \geq \frac{\const{L}}{n}   + \frac{1}{n}  \MIcond{\vect{W}^{  [f]\setminus \theta_{[\gamma],1}  }  }{Q^{(\gamma)}_{[n]} A^{(\gamma)}_{[n]}}{\vect{W}^{\theta_{[\gamma],1}}}.  \label{eq:lem2}
   		\end{IEEEeqnarray}
   	
   	 Now, starting by $\gamma=2$, then applying \eqref{eq:lem2} repeatedly for $\gamma\in [3:\Gamma]$, we have 
    \begin{IEEEeqnarray*}{lCr}
  \MIcond{\vect{W}^{[\theta_{1,2}:f]}}{Q^{(1)}_{[n]} A^{(1)}_{[n]}}{\vect{W}^{(\theta_{1,1})}}\\
  \geq \frac{\const{L}}{n}  + \frac{1}{n}  \MIcond{\vect{W}^{ [f]\setminus \theta_{[2],1}  }}{Q^{(2)}_{[n]} A^{(2)}_{[n]}}{\vect{W}^{\theta_{[2],1} }}  \\
   \geq \frac{\const{L}}{n}  + \frac{1}{n} \left[ \frac{\const{L}}{n}  + \frac{1}{n} \MIcond{\vect{W}^{ [f]\setminus\theta_{[3],1} } }{Q^{(3)}_{[n]} A^{(3)}_{[n]}}{\vect{W}^{ \theta_{[3],1} }} \right] \\ 
    = \frac{\const{L}}{n}  + \frac{\const{L}}{n^2} + \frac{1}{n^2}  \MIcond{\vect{W}^{ [f]\setminus \theta_{[3],1}   }}{Q^{(3)}_{[n]} A^{(3)}_{[n]}}{\vect{W}^{ \theta_{[3],1} }} \\
    \geq \qquad \vdots\\
    \geq \frac{\const{L}}{n} + \dots +\frac{\const{L}}{n^{\Gamma-2}} +\frac{1}{n^{\Gamma-2}} \MIcond{\vect{W}^{ [f]\setminus\theta_{[\Gamma-1],1}  }}{Q^{(\Gamma-1)}_{[n]} A^{(\Gamma-1)}_{[n]}}{\vect{W}^{ \theta_{[\Gamma-1],1} }}\\
     \geq \frac{\const{L}}{n} + \dots +\frac{\const{L}}{n^{\Gamma-2}} + \frac{\const{L}}{n^{\Gamma-1}}  +\frac{1}{n^{\Gamma-1}} \MIcond{\vect{W}^{  [f]\setminus \theta_{[\Gamma],1}  }}{Q^{(\Gamma)}_{[n]} A^{(\Gamma)}_{[n]}}{\vect{W}^{ \theta_{[\Gamma],1} }}\\
    \stackrel{(a)}{=}  \frac{\const{L}}{n} + \dots +\frac{\const{L}}{n^{\Gamma-2}} + \frac{\const{L}}{n^{\Gamma-1}}  +\frac{1}{n^{\Gamma-1}} \underbrace{ \MIcond{   \vect{W}^{[f]\setminus \theta_{[\Gamma],1} } }{ A^{(\Gamma)}_{[n]}}{ Q^{(\Gamma)}_{[n]} \vect{W}^{ \theta_{[\Gamma],1} }} 
     }_{=0}.\\
  \end{IEEEeqnarray*}
 In $(a)$ the last term equals zero due to the independence of the messages and the queries \eqref{eq:IndepQM} and the fact that the answer strings  are a deterministic function of the queries and a \emph{sufficient} number of messages from each classes (see \eqref{eq:DeterministicAnswers},  \eqref{eq:AnswersFromPartialDS}).  Specifically, by combining 
\eqref{eq:DeterministicAnswers} and  \eqref{eq:AnswersFromPartialDS} we have 
\begin{IEEEeqnarray*}{lCr}
	\bigHPcond{A^{(\Gamma)}_{[n]}}{Q^{(\Gamma)}_{[n]} \vect{W}^{ \theta_{[\Gamma],1}  } }\\
	=	\bigHPcond{A^{(\Gamma)}_{[n]}}{Q^{(\Gamma)}_{[n]} \vect{W}^{(\theta_{1,1})}\vect{W}^{ \theta_{[2:\Gamma],1} } } \\
	= \bigHPcond{A^{(\Gamma)}_{[n]}}{Q^{(\Gamma)}_{[n]} \vect{W}^{[\theta_{1,1}:\theta_{2,1}-1]} 	\vect{W}^{(\theta_{2,1})}  \vect{W}^{ \theta_{[3:\Gamma],1}} } \\
	=	\vdots\\
	= \bigHPcond{A^{(\Gamma)}_{[n]}}{Q^{(\Gamma)}_{[n]}\vect{W}^{[f]}}=0.
\end{IEEEeqnarray*}  
    As a result, we obtain 
      \begin{IEEEeqnarray}{rCl} \label{eq:converseIteration}
    \hspace{-4ex} \MIcond{\vect{W}^{[\theta_{1,2}:f]}}{Q^{(1)}_{[n]} A^{(1)}_{[n]}}{\vect{W}^{(\theta_{1,1})}}  \geq \frac{\const{L}}{n} + \dots +\frac{\const{L}}{n^{\Gamma-2}} + \frac{\const{L}}{n^{\Gamma-1}}. 
      \end{IEEEeqnarray}
      Combining \eqref{eq:converseIteration} and \eqref{eq:lem1} yields 
        \begin{IEEEeqnarray}{rCl} 
    \const{L}\left(\frac{1}{\const{R}}-1\right)  \geq \frac{\const{L}}{n} + \dots +\frac{\const{L}}{n^{\Gamma-2}} + \frac{\const{L}}{n^{\Gamma-1}}, 
      \end{IEEEeqnarray}
      and by eliminating $\const{L}$ from both sides, we finally obtain 
       \begin{IEEEeqnarray}{rCl} 
    \const{R}\  && \leq  \inv{\left(1+ \frac{1}{n}+ \frac{1}{n^2}+\dots + \frac{1}{n^{\Gamma-1}} \right)}\\
		&&=  \left(1-\frac{1}{n}\right)  \inv{\left(1-\frac{1}{n^{\Gamma}}\right)}.
      \end{IEEEeqnarray}
\end{proof}
	

\subsection{Achievability of \Cref{thm:MS-PPIR}}\label{sec:MS-PPIRscheme}
	We now present a scheme that achieve the PPIR capacity bound of \cref{thm:MS-PPIR}.
	The capacity of the PIR problem with $n$ noncolluding replicated databases, each storing $f$ messages, was characterized in \cite{SunJafar17_1} as $ \left(1-\frac{1}{n}\right)  \inv{\left(1-\frac{1}{n^{f}}\right)}.$ From the capacity bound of PPIR in  \cref{thm:MS-PPIR} one can observe that  PPIR effectively reduces the size of the database from $f$ to $\Gamma$ messages.  
	 Thus, for our achievable PPIR scheme we  adapt  the capacity achieving PIR scheme  in \cite{SunJafar17_1} to the PPIR problem setup.

	Given $\Gamma$, $n$, $\gamma\in [\Gamma]$, and $\delta= \text{LCM}(M_1,\dots,M_\Gamma)$, the high-level implementation of the PPIR scheme is outlined with the following steps. 
		\begin{enumerate}
			\item The user selects a number $s$ uniformly at random from the set $[\delta]$. 
			\item The user constructs queries $Q^{(\gamma)}_1,\dots, Q^{(\gamma)}_n $  according to \cite[Section IV]{SunJafar17_1}. 
			We assume that the databases store  $\Gamma$  candidate messages $\{\vect{X}^{(1)},\vect{X}^{(2)},\dots, \vect{X}^{(\Gamma)}\}$. Each candidate message is of length $\const{L}=n^{\Gamma}$ \cite{SunJafar17_1} and the user intends to privately retrieve $\vect{X}^{(\gamma)},\,  \gamma\in [\Gamma].$
			\item The user sends the selected random number 
			from Step 1, $s\in [\delta]$, followed by the constructed queries $Q^{(\gamma)}_1,\dots, Q^{(\gamma)}_n, $ in a random order to each database $j\in [n]$. This ensures that if the protocol is applied multiple times with different $s$ and fixed $\gamma$, the user receives a randomized message with probability $1/M_{\gamma}$ for any $\gamma\in [\Gamma]$.
			\item Given the random number $s\in[\delta],$ each database $j\in [n]$ selects a subset of size $\Gamma$ from the messages, one message from each class,  to be used in constructing its answer string $A^{(\gamma)}_j$. 
			The indices of the selected messages  are computed as 	\begin{IEEEeqnarray}{lCr}\label{eq:theta_selection}
			{\theta_{\gamma,\beta_{\gamma,1}}}=\left\lceil \frac{s}{\delta} M_{\gamma} \right\rceil + \sum_{l=1}^{\gamma-1} M_{l},
			\end{IEEEeqnarray} where $\beta_{\gamma,1}= \lceil \frac{s}{\delta} M_{\gamma} \rceil$ is the index of the selected message within its class, and  $\theta_{\gamma,\beta_{\gamma,1}}$ follows from \eqref{eq:indexMapping} due to the fact that the messages are ordered in an ascending order based on their class membership. 
			Each of these $\Gamma$ messages are mapped to the user's queries in Step 2 as $\vect{X}^{(\gamma)}=\vect{W}^{(\theta_{\gamma,\beta_{\gamma,1}})}$ for all $\gamma\in [\Gamma]$. The mapping in \eqref{eq:theta_selection} ensures that each message in a given class is a member of the candidate message set with the same probability.
			\end{enumerate}
		
		\noindent {\bf  \textit{Privacy:}} Note that the query structure of the PIR capacity achieving scheme in \cite{SunJafar17_1} is  fixed independently of the desired \emph{candidate} message index $\gamma\in [\Gamma]$. %
		This fixed structure adheres to three principles to achieve this independence, namely, database symmetry, message symmetry and side-information exploitation \cite{SunJafar17_1}.  
		%
		%
		Moreover, since the achievable construction of \cite{SunJafar17_1} guarantees that any message $\vect{X}^{(\gamma)}$ for all $\gamma\in [\Gamma]$ is equally likely to be the desired message, it follows that the $\Gamma$ messages $\vect{W}^{(\theta_{\gamma,\beta_{\gamma,1}})}$ for $\gamma\in [\Gamma]$, i.e., one message from each class, are also equally likely to be the desired messages. As a result,  ${\mathbb P}(\gamma=\gamma_i| Q^{(\gamma)}_j, A^{(\gamma)}_j )=\frac{1}{\Gamma} $ for any $\gamma_i\in [\Gamma],$ $j\in [n],$ and the query and answer string of any desired class $\gamma\in [\Gamma]$ are indistinguishable from the perspective of each database. This in turns satisfies the M-PPIR privacy constraint in \eqref{eq:MPPIR-privacy}.
		
		\noindent {\bf  \textit{Correctness:}}  Given that the scheme in \cite{SunJafar17_1} guarantees the retrieval of all the $n^{\Gamma}$ symbols of $X^{(\gamma)},$ which is mapped by each database to $\vect{W}^{(\theta_{\gamma,\beta_{\gamma,1}})}$, the user obtains all the symbols of a message that belongs to the class $\gamma$. Thus, the M-PPIR correctness constraint of \eqref{eq:MPPIR-correctness} is satisfied.
		
		\noindent {\bf  \textit{Calculation of the achievable rate:}} For privately retrieving one message from a candidate set of size $\Gamma$ from $n$ replicated databases,  the scheme of \cite{SunJafar17_1} achieves an information retrieval rate of $\inv{\left(1+ \frac{1}{n}+ \frac{1}{n^2}+\dots + \frac{1}{n^{\Gamma-1}} \right)}$, as shown in \cite[Thm.~1]{SunJafar17_1},  which matches the PPIR capacity of \cref{thm:MS-PPIR}.

	The key concepts of the capacity-achieving PPIR scheme are illustrated with the following example.

	\begin{example} Consider the case where we have a number of $f=20$ messages classified into $\Gamma=3$ classes where the number of messages in each class are given by $[4,6,10]$, respectively. The $f$ messages are replicated in $n=2$ databases. Suppose that the user is interested in retrieving a message from class $\gamma=3$. 

\medskip
{\bf  \textit{Step 1 and 2: Queries to databases:}} 
	First, the user selects a number $s\in[\delta]$, where $\delta\triangleq \text{LCM}(4,6,10)=60$,  uniformly at randomly and send this number to the $n$ databases. 
	
	Next, the user utilizes the achievable scheme in \cite{SunJafar17_1} to generate the query sets for privately retrieving one message from a set of $\Gamma$ candidate messages $\{\vect{X}^{(1)},\vect{X}^{(2)},\vect{X}^{(3)}\}$ where $\vect{X}^{(\gamma)}=\{X^{(\gamma )}_1, X^{(\gamma)}_2\dots, X^{(\gamma)}_{\const{L}}\}$, for $\gamma\in [3]$. The query generation steps below precisely follow the steps outlined in \cite{SunJafar17_1} and are presented here for completeness.

	The achievable scheme in \cite{SunJafar17_1} requires the size of each message to be $\const{L}=n^{\Gamma}=8$ and its query sets are constructed as follows. First, to make the symbols downloaded from each database appear random and independent from the desired message, the indices of the $\const{L}$ symbols of each message are  randomly permuted prior to the query construction. Let,  $U^{(\gamma)}_i=X^{(\gamma)}_{\pi_{\gamma}(i)}, \forall i\in[\const{L}], \gamma\in[\Gamma]$, where $\pi_{\gamma}(\cdot)$ is a uniform random permutation privately selected by the user independently for each candidate message. We simplify the notation by letting $U^{(1)}_i=x_i$, $U^{(2)}_i=y_i$ and $U^{(3)}_i=z_i$  for $i\in[\const{L}]$. 
	To retrieve a message from the desired class $\gamma=3$, i.e.,  the candidate message $\vect{z}=\{z_1,z_2,\dots,z_8\}$, symbols are queried from the two databases in a total of $\tau=3$ rounds.  This is shown in \cref{tab:answers-table}(a) where the queries of round $\tau$ are indicated with $Q^{(\gamma)}_j(\tau)$.
	
	{\textit{Initialization Round (${\tau=1}$)}:} The user first queries $(n-1)^{\tau-1}=1$ distinct instance of $z_{i}$ from each database. By message and index symmetries this also applies to $x_{i}$ and $y_{i}$, resulting in total $n{\Gamma \choose 1}(n-1)^{(1-1)}=6$ symbols.  The symbols queried in the first round are shown in the row indicated by $Q^{(3)}_j(1)$ in \cref{tab:answers-table}(a).
	
	{\textit{Following Rounds (${\tau\in[2:3]}$)}:}  In each round and for each database, the user further queries sums of $\tau$ symbols with each symbol is from a different message. The queried sums either contain a single symbol from the desired message (so-called desired $\tau$-sums) or only symbols from undesired messages (so-called undesired $\tau$-sums,  referred to as \emph{side information}). One can see that by utilizing the undesired $\tau$-sums obtained from the previous round, the desired message can be decoded. For example, in round $\tau=3$, the desired symbol $z_{7}$ can be obtained by canceling the side information $x_{6}+y_{5}$ which is obtained from the $2$nd database in round $\tau=2$. Similarly, one can verify the successful recovery of all symbols of the desired message $\vect{z}$ from the queried desired $\tau$-sums shown in \cref{tab:answers-table}(a). Note that after deciding which desired sums to query, the undesired sums to query can be decided by enforcing message and index symmetry and the total number of symbols queried in round $\tau$ is equal to $n{\Gamma \choose \tau}(n-1)^{(\tau-1)}$. Finally, the queries are sent to each database $j\in [2]$.
	%

\begin{table}[h!]
	\centering
	\caption{Query sets for a message from an $n=2$ replication-based DSS storing $f=20$ messages which are classified into $\Gamma=3$ classes. ($\textnormal{a}$) shows the query sets for desired class $\gamma=3$ and ($\textnormal{b}$) for $\gamma=1$, respectively. }
	\label{tab:answers-table}
		
		\begin{minipage}{.4\linewidth}
			\centering
		\begin{IEEEeqnarraybox}[
			\IEEEeqnarraystrutmode
			\IEEEeqnarraystrutsizeadd{4pt}{2pt}]{v/c/v/c/v/c/v}
			\IEEEeqnarrayrulerow\\
			& j && 1 && 2\\
			\hline\hline
			& Q^{(3)}_j(1)
			&& x_{1},\, y_{1},\, z_{1} &&  x_{2},\,  y_{2},\, z_{2}  &
			\\*\cline{1-7}      
			& \multirow{3}{*}{$Q^{(3)}_j(2)$}
			&& x_{4}+y_{3} &&  x_{6}+y_{5}  &
			\\ 
			& 
			&& x_{2}+z_{3} &&  x_{1}+z_{5}  &
			\\
			&
			&& y_{2}+z_{4} &&  y_{1}+z_{6} &
			\\*\cline{1-7}      
			& Q^{(3)}_j(3)
			&& x_{6}+y_{5}+z_{7} && x_{4}+y_{3}+z_{8} &
			\\*\IEEEeqnarrayrulerow
		\end{IEEEeqnarraybox}
	
		(a)
		\\[2mm]
		\end{minipage}%
		\begin{minipage}{.4\linewidth}
		\centering
	\begin{IEEEeqnarraybox}[
		\IEEEeqnarraystrutmode
		\IEEEeqnarraystrutsizeadd{4pt}{2pt}]{v/c/v/c/v/c/v}
		\IEEEeqnarrayrulerow\\
		& j && 1 && 2\\
		\hline\hline
		& Q^{(1)}_j(1)
		&& x_{1},\, y_{1},\, z_{1} &&  x_{2},\,  y_{2},\, z_{2}  &
		\\*\cline{1-7}      
		& \multirow{3}{*}{$Q^{(1)}_j(2)$}
		&& x_{3}+y_{2} &&  x_{5}+y_{1}  &
		\\ 
		& 
		&& x_{4}+z_{2} &&  x_{6}+z_{1}  &
		\\
		&
		&& y_{4}+z_{3} &&  y_{6}+z_{5} &
		\\*\cline{1-7}      
		& Q^{(1)}_j(3)
		&& x_{7}+y_{6}+z_{5} && x_{8}+y_{4}+z_{3} &
		\\*\IEEEeqnarrayrulerow
	\end{IEEEeqnarraybox}

(b)
\\[2mm]
	\end{minipage}%
\end{table}
\medskip
{\bf \textit{Steps 3 to 5: Database answers:}} 
	 Assume that the randomly selected number in Step 1) is given as $s=13$. Accordingly, each database selects the same subset of candidate messages as follows: 
	$\vect{X}^{(1)}=\vect{W}^{(\theta_{1,\beta_{1,1}})}$ , $\vect{X}^{(2)}=\vect{W}^{(\theta_{2,\beta_{2,1}})}$ , and $\vect{X}^{(3)}=\vect{W}^{(\theta_{3,\beta_{3,1}})}$ where $\theta_{1,\beta_{1,1}}=\lceil 0.216 \times 4 \rceil =1 $, $\theta_{2,\beta_{2,1}}=\lceil 0.216 \times 6 \rceil + 4=6 $, and $\theta_{3,\beta_{3,1}}=\lceil 0.216 \times 10 \rceil +10=13 $, respectively. Using this mapping between the identity of the candidate messages and the identity of the stored messages\footnote{ Note that, if we assume the user has knowledge of the size of each class, then $\delta$ is not needed. In this case, an achievable scheme is generated by first randomly selecting one message from each class to construct a set of $\Gamma$ candidate messages. The mapping between the class index and the message index is made locally by the user, and the queries are directly generated as PIR queries with the selected messages identities.}, each database then generates its answer string according to the queries of \cref{tab:answers-table}(a). In other words, the query for $x_i$ is answered by each database with the symbol $W^{(1)}_i$, the query of $y_i$ is answered with the symbol $W^{(6)}_i$, and query of $z_i$ is answered with the symbol $W^{(13)}_i$.

\medskip	
{\bf  \textit{Privacy and correctness of the retrieved message:}} 
	By decoding the downloaded symbols, we obtain the corresponding symbols of the message $\vect{W}^{(13)}$ which is indeed a message from the desired class $\gamma=3$. Moreover, since  the achievable scheme in \cite{SunJafar17_1} follows the symmetry principles, i.e., message, index, and database symmetries within the query sets of each database, the privacy is inherently ensured. 
	Specifically, the achievable scheme in \cite{SunJafar17_1} guarantees the private retrieval of the message $\vect{W}^{(13)}$ among the set $\{\vect{W}^{(1)},\vect{W}^{(6)},\vect{W}^{(13)}\}$ from the perspective of each database.  With each message representing a class $\gamma\in [\Gamma]$, the desired class is also indistinguishable.
	 For example, \Cref{tab:answers-table}(b) illustrates the query sets for desired class $\gamma=1$. 
From Tables~\ref{tab:answers-table}(a) and~\ref{tab:answers-table}(b) one can verify that the index  mapping
\begin{IEEEeqnarray}{rCl}    
	&&	\text{Databases 1:}  \quad (1,2,3,4,5,6,7)	\xrightarrow[]{\gamma=1} \, (1,4,2,3,6,7,5), \label{eq:privacy_map_1}  \\ 
	&&	\text{Databases 2:} \quad  (1,2,3,4,5,6,8) \xrightarrow[]{\gamma=1}  \, (6,2,4,8,1,5,3)  \label{eq:privacy_map_2} 
\end{IEEEeqnarray}
converts the queries for $\gamma=3$ to the queries for $\gamma=1$. To see this mapping, compare   $x_{i_1}+y_{i_2}$ and $x_{\hat{i}_1}+y_{\hat{i}_2}$ from the queries of the first database in Tables~\ref{tab:answers-table}(a) and~\ref{tab:answers-table}(b), respectively. It can be seen that the indices $i_1=4$ and $i_2=3$ of the queries for $\gamma=3$ are converted to the indices $\hat{i}_1=3$ and $\hat{i}_2=2$ of the queries for $\gamma=1$, respectively. Thus, we have the mapping $((i_1,i_2)\rightarrow(\hat{i}_1,\hat{i}_2))=((4,3)\rightarrow(3,2))$. A similar comparison between the remaining queries results in the index and sign mappings of \eqref{eq:privacy_map_1} and \eqref{eq:privacy_map_2}. 
One can similarly verify that there exists a mapping from the queries for $\gamma=3$ to the queries for $\gamma=2$, i.e., $Q^{(3)}_{[2]} \leftrightarrow Q^{(2)}_{[2]}$. Since the permutation  $\pi_{\gamma}(t)$ over these indices  is uniformly and privately selected by the user independently of the desired class index $\gamma$, these queries are equally likely and indistinguishable. 	

\medskip	
{\bf  \textit{Achievable Rate:}} By counting the number of symbols to be downloaded as answer for  the queries in  \cref{tab:answers-table}(a), we obtain the PPIR rate $\const{R}=\frac{8}{14}= \frac{4}{7}= \const{C}_{\text{PPIR}}.$

	\end{example}

\section{Multi-Message Pliable Private Information Retrieval}\label{sec:MPPIR}
 
In this section, we consider the general problem of M-PPIR  as
presented in \Cref{sec:system-model} with $\lambda\geq1, \eta\geq1 $ and derive upper and lower bounds on the M-PPIR rate. 
Recall that, as in the PPIR problem, in M-PPIR the user is \emph{oblivious} to the structure of the database. Hence, we cannot directly utilize existing multi-message PIR solutions for the M-PPIR problem.
To this end, 
we consider  replication-based DSSs and derive upper and lower bounds on the M-PPIR rate. 
As mentioned in \Cref{sec:PPIR}, the single-message PPIR problem is a special case of M-PPIR, thus, the results of \Cref{thm:MS-PPIR} can be recovered by setting $\lambda=1$ and $\eta=1$ in the bounds derived in \Cref{thm:MS-MPPIR}  below. We state our main result for M-PPIR over replicated DSS with \Cref{thm:MS-MPPIR} as follows.

\begin{theorem}	\label{thm:MS-MPPIR}
	Consider a DSS with $n$ noncolluding replicated databases, storing $f$ messages classified into $\Gamma$ classes. For the M-PPIR problem with $\lambda\geq1$ and $\eta\geq 1$,  the maximum achievable M-PPIR rate over all possible M-PPIR protocols, i.e., the M-PPIR capacity $\const{C}_{\textnormal{M-PPIR}}$, is given as
		\begin{IEEEeqnarray*}{c}
		 \underbar{\const{R}} \leq	\const{C}_{\textnormal{M-PPIR}}  \leq \bar{\const{R}}
				\end{IEEEeqnarray*}
			where
	\begin{subnumcases}{}	
				\bar{\const{R}} = \underbar{\const{R}}  =   \inv{\left[ 1+\frac{\Gamma-\eta}{n\eta}\right]}  &  \text{if $\eta \geq \frac{\Gamma}{2}$}, 	\label{eq:ConverseBound-MPPIR1}\\[2mm]
				 \bar{\const{R}} = \inv{\left[ \frac{1-(\frac{1}{n})^{\lfloor \frac{\Gamma}{\eta}\rfloor}}{1-\frac{1}{n}}+ \left( \tfrac{\Gamma}{\eta} -\lfloor\tfrac{\Gamma}{\eta} \rfloor \right) (\tfrac{1}{n})^{\lfloor \frac{\Gamma}{\eta}\rfloor} \right]}  & 	\text{if $\eta \leq \frac{\Gamma}{2}$},	\label{eq:ConverseBound-MPPIR2}\\[2mm]
				 \underbar{\const{R}} =  \frac{\sum_{i=1}^{\eta}  \tau_i \kappa_i^{\Gamma-\eta} \left[ \left( 1+ \frac{1}{\kappa_i}\right)^{\Gamma} - \left( 1+ \frac{1}{\kappa_i}\right)^{\Gamma-\eta}  \right]}{\sum_{i=1}^{\eta}  \tau_i \kappa_i^{\Gamma-\eta} \left[ \left( 1+ \frac{1}{\kappa_i}\right)^{\Gamma} - 1  \right]}& 	\text{if $\eta \leq \frac{\Gamma}{2}$}, \label{eq:ConverseBound-MPPIR3}
	\end{subnumcases}
for  $\kappa_i\triangleq\frac{e^{j2\pi(i-1)/\eta}}{n^{(1/\eta)}-e^{j2\pi(i-1)/\eta}}$. Here,   $\tau_i, \, i\in[\eta]$, is the solution of the  $\eta$ linear equations
\begin{subnumcases}{}	
	\sum_{i=1}^{\eta}  \tau_i \kappa_i^{-\eta} = (n-1)^{\Gamma-\eta},\label{eq:lin1}\\ 
	 \sum_{i=1}^{\eta}  \tau_i \kappa_i^{-k}=0	\quad \text{for } k\in[\eta-1].\label{eq:lin2}
\end{subnumcases}
\end{theorem}

The converse bounds of \Cref{thm:MS-MPPIR} are derived in \Cref{sec:MS-MPPIRconverse-case1} and \Cref{sec:MS-MPPIRconverse-case2}, respectively. The achievability lower bounds in \Cref{thm:MS-MPPIR} are shown in \Cref{sec:MS-MPPIRschems}. The following corollary states that 
if $\frac{\Gamma}{\eta}\in \Naturals$, i.e., the number of classes is divisible by the number of desired classes, then the achievability bound of \eqref{eq:ConverseBound-MPPIR3} matches the upper bound  of \eqref{eq:ConverseBound-MPPIR2}.

\begin{corollary}
	\label{cor:G-MPPIR}
	For the M-PPIR problem from $n> 1$ noncolluding replicated databases where $\eta \leq \frac{\Gamma}{2}, \; \frac{\Gamma}{\eta}\in \Naturals $, the derived upper bound of \eqref{eq:ConverseBound-MPPIR2} is tight, i.e., matches the lower bound of \eqref{eq:ConverseBound-MPPIR3}, and the M-PPIR capacity is given by
		\begin{IEEEeqnarray}{lCr} 
 		\const{C}_{\textnormal{M-PPIR}} =  \left(1-\frac{1}{n}\right)  \inv{\left[1-\left(\frac{1}{n}\right)^{\frac{\Gamma}{\eta}} \right]}.
 		\end{IEEEeqnarray}
 
\end{corollary}

The proof of \cref{cor:G-MPPIR} follows, similarly to the proof of \cite[Cor.~3]{BanawanUlukus18_2}, from the fact that $\nicefrac{\Gamma}{\eta}= \lfloor\nicefrac{\Gamma}{\eta}\rfloor $ and $(1+\nicefrac{1}{\kappa_i})=n^{\nicefrac{1}{\eta}}$ for the bounds of \eqref{eq:ConverseBound-MPPIR2}  and \eqref{eq:ConverseBound-MPPIR3}, respectively. 
\begin{remark}
	 \Cref{thm:MS-MPPIR}  and 	\Cref{cor:G-MPPIR}  yield a simple yet powerful observation.  One can observe that privately retrieving multiple messages $\lambda> 1$ from \emph{multiple} desired classes $\eta>1$, while keeping the identity of the desired classes indices hidden from each database, imposes no penalty on the download rate compared to privately retrieving only \emph{one} message from each of the desired classes. 
	 Moreover, the presented bounds match the M-PIR rates for the case where the user is interested in privately retrieving $\eta$ messages from a dataset consisting of $\Gamma$ messages, i.e., each class contains only one message \cite[Thm.~1, Thm.~2, Cor.~3]{BanawanUlukus18_2}. 
\end{remark}

\begin{remark}
	For the single server M-PPIR problem ($n=1$), 
	the M-PPIR capacity $\const{C}_{\textnormal{M-PPIR}}$ for $\lambda$ arbitrary messages out of $\eta\in [\Gamma]$ desired classes is given by $\const{C}_{\textnormal{M-PPIR}}= \frac{ \eta}{\Gamma}.$ This can be shown by following  \Cref{rem:SingleServerPPIR} and by substituting each of the randomly selected $\Gamma$ messages with \emph{super} messages each consisting of $\lambda$ messages from every class. 
\end{remark}

In the following, we first derive an upper bound for the M-PPIR problem 
by adapting the classical  PIR converse proofs of \cite{BanawanUlukus18_2,SunJafar17_1}  to our pliable setup. 
 The key idea of the converse proof is to select the minimum number of subsets with minimum to no overlap from all possible $\Gamma \choose \eta$ subsets in $\mathfrak{S}$, i.e., the set of all unique subsets of $[\Gamma]$ of size $\eta$, such that ${\bigcup}_{i\in [|\mathfrak{S}|]} \Omega_i =[\Gamma]$.   The intuition behind this selection is the fact that the answer strings of these desired subsets provide sufficient information to construct the answers for the remaining possible subsets in $\mathfrak{S}$ (see \eqref{eq:AnswersFromPartialDS}). This results in dividing the converse proof into two cases. The first {addresses the} case $\eta \geq \frac{\Gamma}{2}$ where there is always some overlap between the possible desired subsets of classes, while for $\eta \leq \frac{\Gamma}{2}$ there exists a number of subsets that do not overlap.

\subsection{Converse proof of \cref{thm:MS-MPPIR} for $\eta\geq \frac{\Gamma}{2}$}\label{sec:MS-MPPIRconverse-case1}
 
 Here, since $\eta\geq\frac{\Gamma}{2}$, for any $\Omega, \Omega' \in  \mathfrak{S}$, such that $\Omega \neq \Omega'$, we have $ \Omega\cap\Omega' \neq \phi$. In other words, there is always some overlap between the possible subsets of desired classes, and the minimum overlap between any two subsets is $2\eta-\Gamma.$
 As a result of this overlap we have the following lemma.

 \begin{lemma}For the M-PPIR problem with $\eta \geq \frac{\Gamma}{2}$, the following bound holds
 		\begin{IEEEeqnarray}{lCr}
 	 	\MIcond{ \vect{W}^{[f]\setminus \theta_{[\eta],[\lambda]}} }{Q^{[\eta]}_{[n]} A^{[\eta]}_{[n]}}{\vect{W}^{ \theta_{[\eta],[\lambda]} }} \geq  \frac{\lambda\const{L}}{n} (\Gamma-\eta).  \label{eq:Remainder_entrpy}
 	 		\end{IEEEeqnarray}
  		Moreover, \eqref{eq:Remainder_entrpy} holds for any set $\Omega\in   \mathfrak{S}$, i.e., 
  			\begin{IEEEeqnarray}{lCr}
  			\MIcond{ \vect{W}^{[f]\setminus\theta_{\Omega,[\lambda]}} }{Q^{\Omega}_{[n]} A^{\Omega}_{[n]}}{\vect{W}^{ \theta_{\Omega,[\lambda]} }} \geq  \frac{\lambda\const{L}}{n} (\Gamma-\eta). \label{eq:Remainder_entrpyG}
  		\end{IEEEeqnarray}
  	\label{lem:RemainderLemma}
 	\end{lemma}
 The proof of \cref{lem:RemainderLemma} follows similar steps as the steps for \cref{lem:lem22} and can be found in Appendix~\ref{app:RemainderLemma}. Now, we are ready to prove the converse for the case $\eta\geq \frac{\Gamma}{2}$. 
 \begin{proof}
 	By combining \cref{lem:lem11} and \cref{lem:RemainderLemma}, we have 
 		\begin{IEEEeqnarray}{lCr}
 		 \eta\lambda \const{L}\left(\frac{1}{\const{R}}-1\right) \geq  \frac{\lambda\const{L}}{n} (\Gamma-\eta),
 	\end{IEEEeqnarray}
and by eliminating $\lambda\const{L}$, we obtain 
 	\begin{IEEEeqnarray}{lCr}
 	\const{R} \leq  \inv{\left[ 1+\frac{\Gamma-\eta}{n\eta}\right]}.
 \end{IEEEeqnarray}
 That proves the upper bound on the M-PPIR capacity for $\eta\geq \frac{\Gamma}{2}$ as given in  \eqref{eq:ConverseBound-MPPIR1}.
 	\end{proof}

\subsection{Converse proof of \cref{thm:MS-MPPIR} for  $\eta\leq \frac{\Gamma}{2}$}\label{sec:MS-MPPIRconverse-case2}

\begin{proof}		
	Without loss of generality, let $\Omega_1=[\eta]$, $\Omega_i =[\eta(i-1)+1: \eta(i)]$ for $i\in [2:\rho]$ and $\rho=\lfloor\frac{\Gamma}{\eta} \rfloor$. Let $\Omega_{\rho'}=[\Gamma-\eta+1:\Gamma]$. We have ${\bigcap}_{i=1}^\rho \Omega_i =\phi$, $\Omega_\rho \cap \Omega_{\rho'}=[\Gamma-\eta+1: \eta \lfloor\frac{\Gamma}{\eta} \rfloor ],$ and $\left\{\ {\bigcup}_{i=1}^\rho \Omega_i \right\} \cup \Omega_{\rho'} =[\Gamma].$ 
	Starting by $\Omega_1=[\eta]$, then applying \cref{lem:lem22} repeatedly we have 
	\begin{IEEEeqnarray}{lCr}
		\MIcond{\vect{W}^{ [f]\setminus{ \theta_{[\eta],[\lambda]}} }}{Q^{\Omega_1}_{[n]} A^{\Omega_1}_{[n]}}{\vect{W}^{ \theta_{[\eta],[\lambda]} }} \nonumber \\
		\geq \frac{\eta\lambda\const{L}}{n}  + \frac{1}{n}  \MIcond{\vect{W}^{ [f]\setminus{ \theta_{[2\eta],[\lambda]}  }  }}{Q^{\Omega_2}_{[n]} A^{\Omega_2}_{[n]}}{\vect{W}^{ \theta_{[2\eta],[\lambda]} }}   \nonumber\\
		\geq \frac{\eta\lambda\const{L}}{n}  + \frac{1}{n} \left[ \frac{\eta\lambda\const{L}}{n}  + \frac{1}{n} \MIcond{\vect{W}^{ [f]\setminus{ \theta_{[3\eta],[\lambda]} } } }{Q^{\Omega_3}_{[n]} A^{\Omega_3}_{[n]}}{\vect{W}^{ \theta_{[3\eta],[\lambda]} }} \right]  \nonumber \\
		= \frac{\eta\lambda\const{L}}{n}  + \frac{\eta\lambda\const{L}}{n^2} + \frac{1}{n^2}  \MIcond{\vect{W}^{[f]\setminus{\theta_{[3\eta],[\lambda]} } }}{Q^{\Omega_3}_{[n]} A^{\Omega_3}_{[n]}}{\vect{W}^{ \theta_{[3\eta],[\lambda]} }}  \nonumber \\
		\geq \qquad \vdots \nonumber \\
		\geq \frac{\eta\lambda\const{L}}{n} + \dots +\frac{\eta\lambda\const{L}}{n^{\lfloor\frac{\Gamma}{\eta}\rfloor-2}} +\frac{1}{n^{\lfloor\frac{\Gamma}{\eta}\rfloor-2}} \MIcond{\vect{W}^{ [f]\setminus{ \theta_{[\eta\lfloor\frac{\Gamma}{\eta}\rfloor-1],[\lambda]} }   }}{Q^{\Omega_{\rho-1}}_{[n]} A^{\Omega_{\rho-1}}_{[n]}}{\vect{W}^{ \theta_{[\eta\lfloor\frac{\Gamma}{\eta}\rfloor-1],[\lambda]} }} \nonumber \\
		\geq \frac{\eta\lambda\const{L}}{n} + \dots +\frac{\eta\lambda\const{L}}{n^{\lfloor\frac{\Gamma}{\eta}\rfloor-2}} + \frac{\eta\lambda\const{L}}{n^{\lfloor\frac{\Gamma}{\eta}\rfloor-1}}  +\frac{1}{n^{\lfloor\frac{\Gamma}{\eta}\rfloor-1}} \MIcond{\vect{W}^{  [f]\setminus{\theta_{[\eta\lfloor\frac{\Gamma}{\eta}\rfloor],[\lambda]} }   }}{Q^{\Omega_\rho}_{[n]} A^{\Omega_\rho}_{[n]}}{\vect{W}^{ \theta_{[\eta\lfloor\frac{\Gamma}{\eta}\rfloor],[\lambda]} }} \nonumber\\
		\geq \frac{\eta\lambda\const{L}}{n} + \dots +\frac{\eta\lambda\const{L}}{n^{\lfloor\frac{\Gamma}{\eta}\rfloor-2}} + \frac{\eta\lambda\const{L}}{n^{\lfloor\frac{\Gamma}{\eta}\rfloor-1}} + \frac{1}{n^{\lfloor\frac{\Gamma}{\eta}\rfloor}} \Big[ \lambda\const{L} \big(\Gamma- \eta\lfloor\tfrac{\Gamma}{\eta}\rfloor \big) \Big],  \label{eq:GconverseIteration}
	\end{IEEEeqnarray}
where \eqref{eq:GconverseIteration} results from bounding the last mutual information term, similar to \cref{lem:RemainderLemma}, as follows
{ 
\begin{IEEEeqnarray*}{rCl} 
n \MIcond{\vect{W}^{  [f]\setminus{\theta_{[\eta\lfloor\frac{\Gamma}{\eta}\rfloor],[\lambda]} }   }}{Q^{\Omega_\rho}_{[n]} A^{\Omega_\rho}_{[n]}}{\vect{W}^{\theta_{[\eta\lfloor\frac{\Gamma}{\eta}\rfloor],[\lambda]} }} 
\geq  \lambda\const{L} \left(\Gamma-\eta \lfloor\tfrac{\Gamma}{\eta}\rfloor\right).
	\end{IEEEeqnarray*}
}
	Now, combining \eqref{eq:GconverseIteration} and \cref{lem:lem11} yields
	\begin{IEEEeqnarray}{rCl} %
		\eta\lambda\const{L}\left(\frac{1}{\const{R}}-1\right)  %
		\geq  \eta\lambda\const{L} \left( \frac{1}{n}  + \dots + 
		 \frac{1}{n^{\lfloor\frac{\Gamma}{\eta}\rfloor-2}} + \frac{1}{n^{\lfloor\frac{\Gamma}{\eta}\rfloor-1}}  + \frac{1}{n^{\lfloor \frac{\Gamma}{\eta}\rfloor}}  \Big[ \tfrac{\Gamma}{\eta}- \lfloor\tfrac{\Gamma}{\eta}\rfloor \Big]  \right).
	\end{IEEEeqnarray}
	Eliminating $\eta\lambda\const{L}$ from both sides, we obtain 
	\begin{IEEEeqnarray}{rCl} %
		\const{R} && \leq  \inv{\left( 1+ \frac{1}{n}+ \frac{1}{n^2}+\dots + \frac{1}{n^{\lfloor\frac{\Gamma}{\eta}\rfloor-1}}  + \frac{1}{n^{\lfloor \frac{\Gamma}{\eta}\rfloor}}  \Big[ \tfrac{\Gamma}{\eta}- \lfloor\tfrac{\Gamma}{\eta}\rfloor \Big]  \right)}\\
		%
		%
		&& = \inv{\left[ \frac{1-(\frac{1}{n})^{\lfloor \frac{\Gamma}{\eta}\rfloor}}{1-\frac{1}{n}}+ \frac{ \tfrac{\Gamma}{\eta} -\lfloor\tfrac{\Gamma}{\eta} \rfloor }{n^{\lfloor \frac{\Gamma}{\eta}\rfloor}} \right]},
	\end{IEEEeqnarray}
 which proves the upper bound on the M-PPIR capacity for the case $\eta\leq  \frac{\Gamma}{2}$ as given in  \eqref{eq:ConverseBound-MPPIR2}.
\end{proof}

\subsection{Achievability of \Cref{thm:MS-MPPIR}}\label{sec:MS-MPPIRschems}

The M-PPIR schemes needed for  \Cref{thm:MS-MPPIR} utilize the single-message and multi-message PIR solutions of \cite{ SunJafar17_1,BanawanUlukus18_2}. If we only consider retrieving a single-message from multiple desired classes, i.e., $\lambda=1$ and $\eta\geq 1$, we can  adapt the multi-message scheme of \cite{BanawanUlukus18_2}, 
	similarly to the approach for PPIR in \Cref{sec:MS-PPIRscheme}.  In the following, we outline the required steps for this adaptation with the extension to 
	multiple desired messages $\lambda\geq 1$. 
 The achievable rate of the M-PIR problem with $n$ noncolluding replicated databases, each storing $f$ messages, and $\lambda\eta$ desired messages to download is characterized in \cite[Thm.~1, Thm.~2]{BanawanUlukus18_2}, as 
 	\begin{subnumcases}{ \underbar{\const{R}}=}	
 	  \inv{\left[ 1+\frac{f-\lambda\eta}{n\lambda\eta}\right]}  &  \text{if $\lambda\eta \geq \frac{f}{2}$}, 	\label{eq:LowerBound-MPIR1}\\[2mm]
 	 \frac{\sum_{i=1}^{\lambda\eta}  \tau_i \kappa_i^{f-\lambda\eta} \left[ \left( 1+ \frac{1}{\kappa_i}\right)^{f} - \left( 1+ \frac{1}{\kappa_i}\right)^{f-\lambda\eta}  \right]}{\sum_{i=1}^{\lambda\eta}  \tau_i \kappa_i^{f-\lambda\eta} \left[ \left( 1+ \frac{1}{\kappa_i}\right)^{f} - 1  \right]}& 	\text{if $\lambda\eta \leq \frac{f}{2}$}, \label{eq:LowerBound-MPIR2}
 	\end{subnumcases}
 where  $\kappa_i\triangleq\frac{e^{j2\pi(i-1)/\lambda\eta}}{n^{(1/\lambda\eta)}-e^{j2\pi(i-1)/\lambda\eta}}$. Here,   $\tau_i, \, i\in[\lambda\eta]$, is the solution of the $\eta\lambda$ linear equations 
 %
 \begin{subnumcases}{}
 	\sum_{i=1}^{\lambda\eta}  \tau_i \kappa_i^{-\lambda\eta} = (n-1)^{f-\lambda\eta},  \\
 	\sum_{i=1}^{\lambda\eta}  \tau_i \kappa_i^{-k}=0 \text{ for } k\in[\lambda\eta-1].
\end{subnumcases}
 
 From comparing the upper bounds of M-PPIR of \Cref{thm:MS-MPPIR} in \eqref{eq:ConverseBound-MPPIR1}-\eqref{eq:ConverseBound-MPPIR2} with \eqref{eq:LowerBound-MPIR1}-\eqref{eq:LowerBound-MPIR2}, we can observe that M-PPIR effectively reduces the size of the database from $f$ to $\Gamma$ messages and the number of desired messages from $\lambda\eta$ to simply $\eta$.  
 %
 Thus, for our achievable M-PPIR schemes, 
 we adapt the M-PIR achievable schemes in \cite{BanawanUlukus18_2} to the M-PPIR problem setup for $\eta\geq \frac{\Gamma}{2}$ and $\eta\leq \frac{\Gamma}{2}$, 
 respectively. 
 
 Given $\Gamma$, $\eta$, $\lambda$, $n$, $\Omega \in  \mathfrak{S}$, and $\delta= \text{LCM}(M_1,\dots,M_\Gamma)$, the high-level implementation of our M-PPIR schemes are outlined with the following steps. 
	\begin{enumerate}
		\item The user selects a number uniformly at random from the set $[\delta]$.  
		\item If $\eta\geq \frac{\Gamma}{2}$, the user constructs the queries $Q^{\Omega}_1,\dots, Q^{\Omega}_n$ according to the achievable M-PIR scheme in \cite[Sec.~IV]{BanawanUlukus18_2}. 
		We assume that the databases store $\Gamma$  candidate \emph{super} messages. 
		Each \emph{super} message is of length $\hat{\const{L}}=n^2$ super symbols, i.e., $\vmat{X}^{(\gamma)}=  (\vect{X}^{(\gamma)}_1, \dots, \vect{X}^{(\gamma)}_{\hat{\const{L}}})$ and the super symbol {\m $\vect{X}^{(\gamma)}_l=(X^{(\gamma)}_{l,1},\dots,X^{(\gamma)}_{l,\lambda} )$} corresponds to a vector of symbols from the $\lambda$ messages of class $\gamma\in [\Gamma]$. The user intends to privately retrieve $\eta$ \emph{super} messages $\vmat{X}^{(\gamma_i)},\forall  \gamma_i\in \Omega$.  
		\item The user sends the selected random number from Step 1, $s\in [\delta]$, then the constructed queries $Q^{\Omega}_1,\dots, Q^{\Omega}_n$ in a random order to each database $j\in [n]$. This ensures that if the protocol is applied multiple times with different $s$ and fixed $\Omega$, the user receives randomized messages, each with probability $1/M_{\gamma_i}$ for all $\gamma_i\in \Omega$.
		\item Given the random number $s\in[\delta],$ each database $j\in [n]$ computes the indices of $\lambda\Gamma$ messages, $\lambda$ messages from each class. These indices are computed as follows:
		\begin{itemize}
	    	\item The first message from each class is given by 
	    		\begin{IEEEeqnarray}{lCr}\label{eq:theta_selection1}
	    		\theta_{\gamma_i,\beta_{\gamma_i,1}} =\left\lceil \frac{s}{\delta} M_{\gamma_i} \right\rceil + \sum_{l=1}^{\gamma_i-1} M_{l},
	    	\end{IEEEeqnarray} 
	    	where $\beta_{\gamma_i,1}= \lceil \frac{s}{\delta} M_{\gamma_i} \rceil$ and $\theta_{\gamma_i,\beta_{\gamma_i,1}}$ in \eqref{eq:theta_selection1} follows due to the fact that the messages are ordered in an ascending order based on their class membership as outlined in \Cref{sec:system-model}. 
	    	\item The following $\lambda-1$ messages from each class are selected, without loss of generality, in a cyclic order\footnote{Note that the the achievable M-PPIR schemes are not unique in terms of the selected $\lambda\Gamma$ \emph{candidate} messages. Particularly, we chose a cyclic order to guarantee the selection of a unique subset of $\lambda$ messages from each class to construct the scheme's answer string.} over the members of the class starting with $ \theta_{\gamma_i,\beta_{\gamma_i,2}}= \theta_{\gamma_i,\beta_{\gamma_i,1}}+1$. That is, for any $k\in [\lambda]$, 
	    	\begin{IEEEeqnarray}{lCr} \label{eq:theta_selection2}
	    		\theta_{\gamma_i,\beta_{\gamma_i,k+1}}=
	    		\begin{cases}{}
	    			\theta_{\gamma_i,\beta_{\gamma_i,k}}-M_{\gamma_i}+1 & \text{if } \theta_{\gamma_i,\beta_{\gamma_i,k}}= \sum_{l=1}^{\gamma_i} M_{l}\\ 
	    			\theta_{\gamma_i,\beta_{\gamma_i,k}}+1 & \text{otherwise.}
	    		\end{cases}
	    	\end{IEEEeqnarray}
	    	
        \end{itemize}
        \item Super messages are assembled in each database using the selected $\lambda\Gamma$ messages of the previous step to be used in constructing its answer string $A^{\Omega}_j$ as follows. Each of  $\Gamma$ \emph{super} messages are mapped to the user's queries of Step 2 as $\vect{X}^{(\gamma)}_l=(W^{(\theta_{\gamma,\beta_{\gamma,1}})}_l \, W^{(\theta_{\gamma,\beta_{\gamma,2}})}_l,\dots, W^{(\theta_{\gamma,\beta_{\gamma,\lambda}})}_l )$ for all $\gamma\in [\Gamma]$ and $l\in \hat{\const{L}} $. Note that any operation involving a super symbol is performed element wise.
        \item If $\eta\leq \frac{\Gamma}{2}$, repeat steps 1-5 by constructing the queries $Q^{\Omega}_1,\dots, Q^{\Omega}_n$  according to the achievable M-PIR scheme in \cite[Sec.V]{BanawanUlukus18_2}. 
        Instead of length $\hat{\const{L}}$ in Step~2, here the length of the $\Gamma$ candidate \emph{super} messages is given by
        $$\hat{\const{L}} = \frac{1}{\eta} \sum_{i=1}^{\eta}  \tau_i \kappa_i^{\Gamma-\eta} \left[ \left( 1+ \frac{1}{\kappa_i}\right)^{\Gamma} - \left( 1+ \frac{1}{\kappa_i}\right)^{\Gamma-\eta}  \right]. $$  
        Here,   $\kappa_i\triangleq\frac{e^{j2\pi(i-1)/\eta}}{n^{(1/\eta)}-e^{j2\pi(i-1)/\eta}}$, with  $\tau_i, \, i\in[\eta]$, being the solutions of the $\eta$ linear equations in \eqref{eq:lin1} and \eqref{eq:lin2}.
	\end{enumerate}
\noindent {\bf\textit{Privacy and Correctness:}} The arguments of privacy and correctness follow from the underlying guarantees of the M-PIR solutions of \cite[Sec. IV]{BanawanUlukus18_2} and \cite[Sec.V]{BanawanUlukus18_2}, similarly to the capacity achieving scheme of PPIR in \Cref{sec:MS-PPIRscheme}. 

\noindent {\bf  \textit{Calculation of achievable rate:}} The achievable rates in \eqref{eq:ConverseBound-MPPIR1} and \eqref{eq:ConverseBound-MPPIR2} follow directly from \eqref{eq:LowerBound-MPIR1} and \eqref{eq:LowerBound-MPIR2} by substituting $f$ with $\Gamma$ and $\lambda\eta$ with $\eta$, respectively. 


\medskip
	The key concepts of the capacity-achieving M-PPIR scheme construction for $\eta\geq \frac{\Gamma}{2}$ are illustrated with the following example.

	\begin{example} Consider the case where we have a number of $f=32$ messages classified into $\Gamma=4$ classes where the number of messages in each class are given by $[4,6,10,12]$, respectively. The $f$ messages are replicated in $n=2$ databases. Suppose that the user is interested in retrieving $\lambda=2$ messages from $\eta=2$ desired classes $\Omega=\{1,3\}$. The steps indicated below refer to the achievable scheme outlined above.
	
	\medskip
	{\bf \textit{Steps 1 and 2: Queries to databases:}}
	First, the user selects a number $s\in[\delta]$, where $\delta\triangleq \text{LCM}(4,6,10,12)=60$,  uniformly at random and sends this number to the $n$ databases. 
	
	Next, the user utilizes the achievable M-PIR scheme in \cite[Sec.~IV]{BanawanUlukus18_2} to generate the query sets for privately retrieving two super messages from a set of $\Gamma$ candidate super messages $\{\vmat{X}^{(1)},\vmat{X}^{(2)},\vmat{X}^{(3)},\vmat{X}^{(4)}\}$ where $\vmat{X}^{(\gamma)}=\{\vect{X}^{(\gamma )}_1, \vect{X}^{(\gamma)}_2\dots, \vect{X}^{(\gamma)}_{\hat{\const{L}}}\}$ for $\gamma\in [4]$ and $\vect{X}^{(\gamma)}_l=\{X^{(\gamma )}_{l,1},\dots, X^{(\gamma)}_{l,\lambda}\}$ for all $l\in [\hat{\const{L}}]$. The query generation steps below precisely follow the steps outlined in \cite[Sec.~IV]{BanawanUlukus18_2} and are presented here for completeness.
	
	The achievable scheme in \cite[Sec.~IV]{BanawanUlukus18_2} requires the size of each super message to be $\hat{\const{L}}=n^{2}=4$ and its query sets are constructed as follows. 
	First, to make the symbols downloaded from each database appear random and independent from the desired class subset $\Omega$, the indices of the $\hat{\const{L}}$ symbols of each super message are  randomly permuted prior to the query construction. Let,  $\vect{U}^{(\gamma)}_i=\vect{X}^{(\gamma)}_{\pi_{\gamma}(i)}, \forall i\in[\hat{\const{L}}], \gamma\in[\Gamma]$, where $\pi_{\gamma}(\cdot)$ is a uniform random permutation privately selected by the user independently for each candidate super message. We simplify the notation by letting $\vect{U}^{(1)}_i=\vect{x}_i$, $\vect{U}^{(2)}_i=\vect{y}_i$, $\vect{U}^{(3)}_i=\vect{z}_i$ and $\vect{U}^{(4)}_i=\vect{w}_i$ for $i\in[\hat{\const{L}}]$. 
	To retrieve $\lambda=2$ messages from the desired classes $\Omega=\{1,3\}$, i.e.,  the candidate super messages $\vmat{x}=\{\vect{x}_1,\vect{x}_2,\dots,\vect{x}_4\}$ and $\vmat{z}=\{\vect{z}_1,\vect{z}_2,\dots,\vect{z}_4\}$, super symbols are queried from the two databases in two rounds, i.e., $\tau=2$.  This is shown in \cref{tab:answers-tableEX5} where the queries of round $\tau$ are indicated with $Q^{\Omega}_j(\tau)$.
	
	\textit{Initialization Round (${\tau=1}$)}: The user first queries one distinct instance of $\vect{x}_{i}$ and $\vect{z}_{i}$ from each database. By message and index symmetries this also applies to $\vect{y}_{i}$ and $\vect{w}_{i}$, resulting in total $n{\Gamma \choose 1}=8$ super symbols.  The queried super symbols in the first round are shown in the row indicated by $Q^{\Omega}_j(1)$ in \cref{tab:answers-tableEX5}.

	\textit{Following Round (${\tau=2}$)}:  In the second round and for each database, the user downloads a linearly encoded mixture of new super symbols from the desired super messages and the super symbols of undesired super messages, i.e., side-information, that are obtained from the other databases in the previous round. Specifically, 
we consider a $[4,2]$ Reed Solomon code $\code{C}$ over $\Field_5$ with  generator matrix
  $\mat{G}^{\code{C}}= \bigl(\begin{smallmatrix} 1 & 1 & 1 & 1\\ 0 & 1 & 2 & 3
  \end{smallmatrix}\bigr).$ The user picks uniformly and independently at random $n-1$ permutations for the columns of $\mat{G}^{\code{C}}$ from the $24$ possible permutations. Each permutation is to be used to encode new desired super symbols with the side information obtained from one of the $n-1$ remaining databases. 
  Here, we have $n=2$. Consider that the user picked the permutation $\{1,3,2,4\}$. Thus, in the second round the queried encoded super symbols for the first database are given by 
   \begin{IEEEeqnarray*}{c}
  \mat{G}^{\code{C}} \begin{pmatrix} 1 & 0 & 0 & 0\\ 0 & 0 & 1 & 0\\ 0 & 1 & 0 & 0\\ 0 & 0 & 0 & 1
  \end{pmatrix} \begin{pmatrix} \vect{x}_3\\ \vect{y}_2\\ \vect{z}_3\\ \vect{w}_2 
  \end{pmatrix}= \begin{pmatrix} 1 & 1 & 1 & 1\\ 0 & 2 & 1 & 3
  \end{pmatrix}\begin{pmatrix} \vect{x}_3\\ \vect{y}_2\\ \vect{z}_3\\ \vect{w}_2 
  \end{pmatrix} =\begin{pmatrix} \vect{x}_3+\vect{y}_2+\vect{z}_3+\vect{w}_2\\ 2\vect{y}_2+\vect{z}_3+3\vect{w}_2  \end{pmatrix}.
  \end{IEEEeqnarray*}
 
    For decodability, one can see that the desired super symbols $\vect{x}_{3}$ and $\vect{z}_{3}$ can be obtained by canceling the side information $\vect{y}_{2}$ and $\vect{w}_{2}$, which are obtained from the $2$nd database in the first round. Similarly, one can verify the successful recovery of all super symbols of the desired super messages $\vmat{x}$ and $\vmat{z}$ from the queried MDS encoded mixtures in \cref{tab:answers-tableEX5}. Finally, the queries are sent to each database $j\in [2]$.
	%
\begin{table}[h!]
	\centering
	\caption{M-PPIR query sets for a desired class set $\Omega=\{1,3\}$ from an $n=2$ replication-based DSS storing $f=32$ messages which are classified into $\Gamma=4$ classes. 
	}
	\label{tab:answers-tableEX5}
		
		\begin{minipage}{.4\linewidth}
			\centering
		\begin{IEEEeqnarraybox}[
			\IEEEeqnarraystrutmode
			\IEEEeqnarraystrutsizeadd{4pt}{2pt}]{v/c/v/c/v/c/v}
			\IEEEeqnarrayrulerow\\
			& j && 1 && 2\\
			\hline\hline
			& Q^{\Omega}_j(1)
			&& \vect{x}_{1},\, \vect{y}_{1},\, \vect{z}_{1},\, \vect{w}_{1} &&  \vect{x}_{2},\,  \vect{y}_{2},\, \vect{z}_{2}  ,\, \vect{w}_{2}&
			\\*\cline{1-7}      
			& \multirow{2}{*}{$Q^{\Omega}_j(2)$}
			&& \vect{x}_3+\vect{y}_2+\vect{z}_3+\vect{w}_2 &&  \vect{x}_4+\vect{y}_1+\vect{z}_4+\vect{w}_1  &
			\\
			&
			&& 2\vect{y}_2+\vect{z}_3+3\vect{w}_2 &&  2\vect{y}_1+\vect{z}_4+3\vect{w}_1 &
			\\*\IEEEeqnarrayrulerow
		\end{IEEEeqnarraybox}
	\end{minipage}%
\end{table}

\medskip
\textbf{\textit{Steps 3 to 5: Database answers:}}
	 Assume that the randomly selected number in Step 1) is given as $s=13$. Accordingly, each database selects the same subset of candidate super messages as follows:
	 $\vect{X}^{(1)}_l=(W^{(\theta_{1,\beta_{1,1}})}_l \, W^{(\theta_{1,\beta_{1,2}})}_l)$, $\vect{X}^{(2)}_l=(W^{(\theta_{2,\beta_{2,1}})}_l \, W^{(\theta_{1,\beta_{2,2}})}_l)$,  $\vect{X}^{(3)}_l=(W^{(\theta_{3,\beta_{3,1}})}_l \, W^{(\theta_{1,\beta_{3,2}})}_l)$, and $\vect{X}^{(4)}_l=(W^{(\theta_{4,\beta_{4,1}})}_l \, W^{(\theta_{4,\beta_{4,2}})}_l)$ for all $l\in [\hat{\const{L}}]$.  
	 By invoking the random selection of $s$ in  \eqref{eq:theta_selection1}, we have   $\theta_{1,\beta_{1,1}}=\lceil 0.216 \times 4 \rceil =1 $, and by \eqref{eq:theta_selection2}  $\theta_{1,\beta_{1,2}}=2$. 
	 Similarly, 
	 $\theta_{2,\beta_{2,1}}=\lceil 0.216 \times 6 \rceil + 4=6 $,  $\theta_{2,\beta_{2,2}}=7$, $\theta_{3,\beta_{3,1}}=\lceil 0.216 \times 10 \rceil +10=13 $, $\theta_{3,\beta_{3,2}}=14$,  $\theta_{4,\beta_{4,1}}=\lceil 0.216 \times 12 \rceil +20=23$, and $\theta_{4,\beta_{4,2}}=24$, respectively. 
	 Using this mapping between the identity of the candidate super messages and the identity of the stored messages, each database then generates its answer string according to the queries of \cref{tab:answers-tableEX5}. 
	 In other words, the query for $\vect{x}_i$ is answered by each database with the symbols $\{W^{(1)}_i, W^{(2)}_i\}$, the query of $\vect{y}_i$ is answered with the symbols $\{W^{(6)}_i,W^{(7)}_i\}$, and so on. 
	
	\medskip
	{\bf  \textit{Privacy and correctness of the retrieved messages:}} 
	By decoding the downloaded symbols, we obtain the corresponding symbols of the messages $\{ \vect{W}^{(1)}, \vect{W}^{(2)}, \vect{W}^{(13)}, \vect{W}^{(14)} \}$ which are indeed $\lambda=2$ messages from each of the desired classes in the set $\Omega=\{1,3\}$.

	Moreover, since  the achievable scheme in \cite[Sec.~IV]{BanawanUlukus18_2} follows the symmetry principles, i.e., message, index, and database symmetries within the query sets of each database in the first round, the privacy is inherently ensured. 
	For the second round, due to the private permutation over the columns of the MDS code, the MDS-encoded mixtures of desired and undesired super symbols do not leak information about the desired set of super messages and thus the desired set of classes. 
	Note that our proposed achievable scheme utilizes super messages instead of simply repeating the construction of \cite[Sec.~IV]{BanawanUlukus18_2} with a randomly selected candidate message set to enforce the selection of unique messages from each class. However, from the privacy perspective, we sequentially utilize $\lambda$ iterations of a perfectly private M-PIR scheme.  
	Specifically, due to the element-wise operations in the second round, the achievable scheme in \cite[Sec.~IV]{BanawanUlukus18_2} guarantees the private retrieval of messages $\{\vect{W}^{(1)}, \vect{W}^{(13)}\}$ from the candidate set $\{\vect{W}^{(1)}, \vect{W}^{(6)}, \vect{W}^{(13)}, \vect{W}^{(23)}\}$ 
	and $\{\vect{W}^{(2)}, \vect{W}^{(14)}\}$ from the candidate set 
	$\{ \vect{W}^{(2)}, \vect{W}^{(7)}, \vect{W}^{(14)}, \vect{W}^{(24)}\}$, resp., 
	from the perspective of each database.  With one  message representing each of the classes $\gamma\in [\Gamma]$ in the two candidate message sets, the desired set of classes is indistinguishable.

\medskip
 {\bf  \textit{Achievable Rate:}} By counting the number of symbols to be downloaded as answer for  the queries in  \cref{tab:answers-tableEX5}, we obtain the M-PPIR rate $\const{R}=\frac{4\times2\times2}{12\times2}= \frac{2}{3}= \const{C}_{\text{M-PPIR}}$ for $\eta \geq \frac{\Gamma}{2}$ as given in \eqref{eq:ConverseBound-MPPIR1}. 
	\end{example}
 
\section{Conclusion}\label{sec:MPPIR-conclusion}

 In this work, we proposed the problem of M-PPIR from noncolluding replicated database as a new variant of the classical PIR problem. 
 In M-PPIR, $f$ messages are 
 classified into $\Gamma$ classes, and the user wishes to retrieve \emph{any} $\lambda\geq 1$ messages from \emph{multiple} desired classes while revealing no information about the identity of the desired classes to the databases. From this general problem, we considered the special case of (single-message) PPIR where the user is interested in retrieving only one message from one desired class. We characterized the PPIR capacity  from replicated noncolluding databases for an arbitrary number of databases $n>1$ and presented capacity-achieving schemes. 
 Interestingly, the capacity of PPIR 
 matches the PIR capacity with $n$ databases and $\Gamma$ messages. However, as a significant contrast to PIR, the answers are randomized if the user queries the same class repetitively. Thus, enabling flexibility, i.e., pliability, allows to trade-off privacy versus download rate compared to classical information-theoretic PIR schemes.
 Moreover, we extended our results to the general M-PPIR problem, derived upper and lower bounds on the M-PPIR rate, and showed a similar insight, i.e., that the derived M-PPIR bounds match the multi-message PIR bounds.  


\appendices

\section{Proof of Lemma \ref{lem:lem11}}\label{app:lem1} 
In this appendix, we prove an upper bound on the conditional mutual information stated in  \cref{lem:lem11}. 
We start the proof of the simplest case\footnote{Note that, for the special case of ($\lambda=1$, $\eta=1$), the proof technique is similar to \cite[Lem~5]{SunJafar17_1}. However, for completeness we restate these steps here using the notation employed in this paper.} where $\lambda=1$ and $\eta=1$ 
as a special case of \cref{lem:lem11}. Then we extend the proof to $\lambda\geq 1$ and $\eta\geq1$. 

	\begin{proof} For $\lambda=1$ and $\eta=1$, we have
		\begin{IEEEeqnarray*}{rCl}
			&&\MIcond{\vect{W}^{[\theta_{1,2}:f]}}{Q^{(1)}_{[n]} A^{(1)}_{[n]}}{\vect{W}^{(\theta_{1,1} )}}\\
			&& \stackrel{(a)}{=}\MI{\vect{W}^{[\theta_{1,2}:f]}}{Q^{(1)}_{[n]} A^{(1)}_{[n]}\vect{W}^{(\theta_{1,1})}} \\
			&& \stackrel{(b)}{=} \MI{\vect{W}^{[\theta_{1,2}:f]}}{Q^{(1)}_{[n]} A^{(1)}_{[n]}} + \underbrace{\MIcond{\vect{W}^{[\theta_{1,2}:f]}}{\vect{W}^{(\theta_{1,1})}}{Q^{(1)}_{[n]}A^{(1)}_{[n]}}}_{=0} \\
			&& \stackrel{(c)}{=} \MIcond{\vect{W}^{[\theta_{1,2}:f]}}{A^{(1)}_{[n]}}{Q^{(1)}_{[n]}} + \underbrace{ \MI{\vect{W}^{[\theta_{1,2}:f]}}{Q^{(1)}_{[n]}}  }_{=0} \\
			&& = \bigHPcond{A^{(1)}_{[n]}}{Q^{(1)}_{[n]}} - \bigHPcond{A^{(1)}_{[n]}}{Q^{(1)}_{[n]}\vect{W}^{[\theta_{1,2}:f]}}\\
			&& \stackrel{(d)}{\leq} \bigHP{A^{(1)}_{[n]}} - \bigHPcond{\vect{W}^{(\theta_{1,1})}A^{(1)}_{[n]}}{Q^{(1)}_{[n]}\vect{W}^{[\theta_{1,2}:f]}}
			+\underbrace{\bigHPcond{\vect{W}^{(\theta_{1,1})}}{A^{(1)}_{[n]}Q^{(1)}_{[n]}\vect{W}^{[\theta_{1,2}:f]}}}_{=0}\\
			&& \stackrel{(e)}{\leq} \const{D} - \bigHPcond{\vect{W}^{(\theta_{1,1})}A^{(1)}_{[n]}}{Q^{(1)}_{[n]}\vect{W}^{[\theta_{1,2}:f]}}\\
			&& \stackrel{(f)}{=} \frac{\const{L}}{\const{R}}- \bigHPcond{\vect{W}^{(\theta_{1,1})}}{Q^{(1)}_{[n]}\vect{W}^{[\theta_{1,2}:f]}}  - \underbrace{\bigHPcond{A^{(1)}_{[n]}}{Q^{(1)}_{[n]}\vect{W}^{(\theta_{1,1})}\vect{W}^{[\theta_{1,2}:f]}}}_{=0} \\
			&& = \frac{\const{L}}{\const{R}}- \const{L}= \const{L}\left(\frac{1}{\const{R}}-1\right)
		\end{IEEEeqnarray*}
		where 
		\begin{itemize}
				\item[$(a)$]  follows from the independence of the messages \eqref{eq:JointEntropy} and the independence of the messages and the queries \eqref{eq:IndepQM}; 
			\item[$(b)$] follows from the chain rule of mutual information and the independence of the messages \eqref{eq:JointEntropy};
			\item[$(c)$] follows from the independence between the messages and the queries \eqref{eq:IndepQM};
			\item[$(d)$] follows from the fact that conditioning reduces entropy \eqref{eq:JointEntropy}, and the correctness condition  \eqref{eq:MPPIR-correctness};
			\item[$(e)$] follows from the chain rule of entropy and \cref{def:def_PCrate};
			\item[$(f)$] follows fact that the answer strings  are a deterministic function of the queries the stored messages  \eqref{eq:DeterministicAnswers}. 
		\end{itemize}

Next, we extend the argument for $\lambda\geq 1$ and $\eta\geq 1$ as follows. Recall that 
 $$\vect{W}^{\theta_{[\eta],[\lambda]}} \triangleq \{ \vect{W}^{(\theta_{1,1})}, \vect{W}^{(\theta_{1,2})},\dots, \vect{W}^{(\theta_{1,\lambda})}, \vect{W}^{(\theta_{2,1})},\dots,\vect{W}^{(\theta_{\eta,1})},\dots,\vect{W}^{(\theta_{\eta,\lambda})} \},$$
 and
 $${\vect{W}^{[f]\setminus{ \theta_{[\eta],[\lambda]} }} \triangleq  \vect{W}^{[\theta_{1,\lambda+1}:\theta_{2,1}-1]}\cup\vect{W}^{[\theta_{2,\lambda+1}:\theta_{3,1}-1]}\cup\dots\cup \vect{W}^{[\theta_{\eta-1,\lambda+1}:\theta_{\eta,1}-1]}\cup  \vect{W}^{[\theta_{\eta,\lambda+1}:f]}.}$$ 
	\begin{IEEEeqnarray*}{rCl}
		&&\MIcond{ \vect{W}^{ [f]\setminus{ \theta_{[\eta],[\lambda]} }}  }{Q^{[\eta]}_{[n]} A^{[\eta]}_{[n]}}{ \vect{W}^{ \theta_{[\eta],[\lambda]}  } }\\
		&& \stackrel{(a)}{=} 
		 \MI{\vect{W}^{ [f]\setminus{ \theta_{[\eta],[\lambda]} }}  }{ Q^{[\eta]}_{[n]} A^{[\eta]}_{[n]} \vect{W}^{ \theta_{[\eta],[\lambda]}  } } \\  %
		&& \stackrel{(b)}{=}
			 \MI{ \vect{W}^{ [f]\setminus{ \theta_{[\eta],[\lambda]} }}  }{ Q^{[\eta]}_{[n]} A^{[\eta]}_{[n]}  }  %
		+  \underbrace{ \MIcond{ \vect{W}^{ [f]\setminus \theta_{[\eta],[\lambda]} }  }{ \vect{W}^{\theta_{[\eta],[\lambda]}  } }{ Q^{[\eta]}_{[n]} A^{[\eta]}_{[n]} }  }_{=0}\\
		&& \stackrel{(c)}{=} \MIcond{\vect{W}^{ [f]\setminus\theta_{[\eta],[\lambda]} } }{ A^{[\eta]}_{[n]}  }{ Q^{[\eta]}_{[n]} } +  \underbrace{ \MI{ \vect{W}^{ [f]\setminus \theta_{[\eta],[\lambda]} }  }{ Q^{[\eta]}_{[n]} }  }_{=0}\\ \\
		&& = \bigHPcond{A^{[\eta]}_{[n]}}{Q^{[\eta]}_{[n]}} - \bigHPcond{A^{[\eta]}_{[n]}}{Q^{[\eta]}_{[n]}  \vect{W}^{ [f]\setminus \theta_{[\eta],[\lambda]}  }  }\\
		&& \stackrel{(d)}{\leq} \bigHP{A^{[\eta]}_{[n]}} - \bigHPcond{ \vect{W}^{ \theta_{[\eta],[\lambda]}  } A^{[\eta]}_{[n]}}{Q^{[\eta]}_{[n]}  \vect{W}^{ [f]\setminus \theta_{[\eta],[\lambda]} }  }
		+\underbrace{\bigHPcond{ \vect{W}^{ \theta_{[\eta],[\lambda]}   } }{A^{[\eta]}_{[n]}Q^{[\eta]}_{[n]} \vect{W}^{ [f]\setminus\theta_{[\eta],[\lambda]} }}  }_{=0}\\
		&& \stackrel{(e)}{\leq} \const{D} - \bigHPcond{ \vect{W}^{\theta_{[\eta],[\lambda]} } A^{[\eta]}_{[n]}}{Q^{[\eta]}_{[n]}  \vect{W}^{ [f]\setminus\theta_{[\eta],[\lambda]}  }  }\\  
		&& \stackrel{(f)}{=} \frac{\eta\lambda\const{L}}{\const{R}} - \bigHPcond{ \vect{W}^{\theta_{[\eta],[\lambda]} }}{Q^{[\eta]}_{[n]}  \vect{W}^{ [f]\setminus\theta_{[\eta],[\lambda]} } }  - \underbrace{ \bigHPcond{A^{[\eta]}_{[n]}}{Q^{[\eta]}_{[n]}  \vect{W}^{\theta_{[\eta],[\lambda]} }  \vect{W}^{ [f]\setminus \theta_{[\eta],[\lambda]}  } 
		 } }_{=0} \\
		&& = \frac{\eta\lambda\const{L}}{\const{R}}- \eta\lambda\const{L}= \eta\lambda\const{L}\left(\frac{1}{\const{R}}-1\right)
	\end{IEEEeqnarray*}
where 
	\begin{itemize}
	\item[$(a)$] follows from the independence of the messages \eqref{eq:JointEntropy} and the independence of the messages and the queries \eqref{eq:IndepQM}; 
	
	\item[$(b)$] follows from the chain rule of mutual information and the independence of the messages \eqref{eq:JointEntropy};
	\item[$(c)$] follows from the independence of the queries and the messages \eqref{eq:IndepQM};
	\item[$(d)$] follows from the fact that conditioning reduces entropy \eqref{eq:JointEntropy}, and from the correctness condition \eqref{eq:MPPIR-correctness};
	\item[$(e)$] follows from the chain rule of entropy and \cref{def:def_PCrate};
	\item[$(f)$] follows the fact that the answer strings  are a deterministic function of the queries and the stored messages  \eqref{eq:DeterministicAnswers}.
\end{itemize}
\end{proof}


\section{Proof of Lemma~ \ref{lem:lem22}}\label{app:lem2}

Here, we prove a lower bound on the conditional mutual information stated in \cref{lem:lem22}. 
We first proof of the simplest case for $\lambda=1$ and $\eta=1$,  then extend the same argument for $\lambda\geq 1$ and $\eta\geq1$.
Before starting the proof, recall that for $\gamma\in[\Gamma]$ we have  $\vect{W}^{\theta_{[\gamma],1}} \triangleq \{ \vect{W}^{(\theta_{1,1})}, \vect{W}^{(\theta_{2,1})},\dots, \vect{W}^{(\theta_{\gamma,1})} \}$.
	\begin{proof} Let $\lambda=1$, $\eta=1$, and $\gamma\in [2:\Gamma]$. We have the following chain of inequalities:
		\begin{IEEEeqnarray*}{lCr}
			n \MIcond{ \vect{W}^{[f]\setminus \theta_{[\gamma-1],1}  }  }{Q^{(\gamma-1)}_{[n]} A^{(\gamma-1)}_{[n]}}{ \vect{W}^{\theta_{[\gamma-1],1} } }\\
			%
			%
			\geq  \sum_{j=1}^{n} \MIcond{ \vect{W}^{[f]\setminus\theta_{[\gamma-1],1} }  }{Q^{(\gamma-1)}_{j} A^{(\gamma-1)}_{j}}{\vect{W}^{ \theta_{[\gamma-1],1} }}\\ 
			\stackrel{(a)}{=} \sum_{j=1}^{n} \MIcond{ \vect{W}^{[f]\setminus\theta_{[\gamma-1],1}  } }{Q^{(\gamma)}_{j} A^{(\gamma)}_{j}}{\vect{W}^{ \theta_{[\gamma-1],1} }}\\ 
			\stackrel{(b)}{=} \sum_{j=1}^{n} \MIcond{\vect{W}^{[f]\setminus\theta_{[\gamma-1],1}   }  }{ A^{(\gamma)}_{j}}{Q^{(\gamma)}_{j}\vect{W}^{\theta_{[\gamma-1],1} }}\\
			\stackrel{(c)}{=} \sum_{j=1}^{n} \bigHPcond{ A^{(\gamma)}_{j}}{Q^{(\gamma)}_{j}\vect{W}^{\theta_{[\gamma-1],1}} } 
			-\underbrace{\bigHPcond{ A^{(\gamma)}_{j}}{Q^{(\gamma)}_{j}\vect{W}^{ \theta_{[\gamma-1],1} }   \vect{W}^{[f]\setminus \theta_{[\gamma-1],1}  } } }_{=0}\\
			\geq \sum_{j=1}^{n} \bigHPcond{ A^{(\gamma)}_{j}}{Q^{(\gamma)}_{[n]}A^{(\gamma)}_{[j-1]}\vect{W}^{ \theta_{[\gamma-1],1} }} \\ 
			\stackrel{(c)}{=}  \sum_{j=1}^{n}  \MIcond{ \vect{W}^{[f]\setminus \theta_{[\gamma-1],1} }  }{ A^{(\gamma)}_{j}}{Q^{(\gamma)}_{[n]}A^{(\gamma)}_{[j-1]}\vect{W}^{ \theta_{[\gamma-1],1} }} \\
			=  \MIcond{ \vect{W}^{[f]\setminus \theta_{[\gamma-1],1} }  }{ A^{(\gamma)}_{[n]}}{Q^{(\gamma)}_{[n]}\vect{W}^{ \theta_{[\gamma-1],1} }} \\
			\stackrel{(b)}{=}  \MIcond{ \vect{W}^{[f]\setminus \theta_{[\gamma-1],1}   } }{ A^{(\gamma)}_{[n]}Q^{(\gamma)}_{[n]}}{\vect{W}^{ \theta_{[\gamma-1],1} }} \\
			\stackrel{(d)}{=}    \MIcond{ \vect{W}^{[f]\setminus \theta_{[\gamma-1],1} }  }{ A^{(\gamma)}_{[n]}Q^{(\gamma)}_{[n]} \vect{W}^{(\theta_{\gamma,1})}}{\vect{W}^{ \theta_{[\gamma-1],1} }} 
			-\underbrace{ \MIcond{ \vect{W}^{[f]\setminus \theta_{[\gamma-1],1} }  }{\vect{W}^{(\theta_{\gamma,1})}} {A^{(\gamma)}_{[n]}Q^{(\gamma)}_{[n]} \vect{W}^{\theta_{[\gamma-1],1} } } }_{=0}\\
			=    \MIcond{\vect{W}^{[f]\setminus \theta_{[\gamma-1],1}  }  }{ \vect{W}^{(\theta_{\gamma,1})}}{\vect{W}^{ \theta_{[\gamma-1],1} }} +  \MIcond{\vect{W}^{[f]\setminus\theta_{[\gamma-1],1} }  }{ A^{(\gamma)}_{[n]}Q^{(\gamma)}_{[n]}}{\vect{W}^{\theta_{[\gamma-1],1}}\vect{W}^{(\theta_{\gamma,1})}}\\
			\stackrel{(e)}{=}   \bigHP{ \vect{W}^{(\theta_{\gamma,1})}} +  \MIcond{\vect{W}^{[f]\setminus \theta_{[\gamma],1}  }  }{ A^{(\gamma)}_{[n]} Q^{(\gamma)}_{[n]} }{\vect{W}^{\theta_{[\gamma-1],1} }\vect{W}^{(\theta_{\gamma,1})}}\\
			\stackrel{(f)}{=}   \const{L}+  \MIcond{\vect{W}^{[f]\setminus\theta_{[\gamma],1} }  }{ A^{(\gamma)}_{[n]}Q^{(\gamma)}_{[n]}}{\vect{W}^{\theta_{[\gamma],1}}}, 
		\end{IEEEeqnarray*}
		where  	
		\begin{itemize}
			\item[$(a)$] follows from the privacy constraint \eqref{eq:MPPIR-privacy};
			\item[$(b)$] follows from the independence between the messages \eqref{eq:JointEntropy} and the  independence between the messages and the queries  \eqref{eq:IndepQM};
			\item[$(c)$] it follows from the fact that the answer strings are a deterministic function of the queries and  the stored messages \eqref{eq:DeterministicAnswers}; 
			\item[$(d)$] follows from the chain rule of mutual information, the independence of the messages \eqref{eq:JointEntropy}, and the correctness condition \eqref{eq:MPPIR-correctness}, in particular from $ 	\bigHPcond{ \vect{W}^{ (\theta_{\gamma,1})  }}{Q^{(\gamma)}_{[n]} A^{(\gamma)}_{[n]} }=0  $;  
			\item[$(e)$] follows from the independence of the messages \eqref{eq:JointEntropy};
			\item[$(f)$] follows from the chain rule of mutual information and the fact that each message consists of $\const{L}$ independent and identically distributed symbols \eqref{eq:Entropy}. 
		\end{itemize} 
	
	Next, we extend the argument for $\lambda\geq 1$ and $\eta\geq 1$ as follows. Recall that 
	$$\vect{W}^{\theta_{[\eta],[\lambda]}} \triangleq \{ \vect{W}^{(\theta_{1,1})}, \vect{W}^{(\theta_{1,2})},\dots, \vect{W}^{(\theta_{1,\lambda})}, \vect{W}^{(\theta_{2,1})},\dots,\vect{W}^{(\theta_{\eta,1})},\dots,\vect{W}^{(\theta_{\eta,\lambda})} \},$$
	and
	$${ \vect{W}^{[f]\setminus \theta_{[\eta],[\lambda]} } \triangleq  \vect{W}^{[\theta_{1,\lambda+1}:\theta_{2,1}-1]}\cup \vect{W}^{[\theta_{2,\lambda+1}:\theta_{3,1}-1]}\cup \dots \cup \vect{W}^{[\theta_{\eta-1,\lambda+1}:\theta_{\eta,1}-1]}\cup \vect{W}^{[\theta_{\eta,\lambda+1}:f]}.}$$ 
		
	Let $\Omega_1, \Omega_2 \in  \mathfrak{S}$, such that $\Omega_1 \cap \Omega_2=\phi$, and without loss of generality assume that $\Omega_1=[\eta]$ and $\Omega_2=[\eta+1:2\eta]$. Then
		\begin{IEEEeqnarray*}{lCr}
				n\MIcond{   \vect{W}^{[f]\setminus\theta_{[\eta],[\lambda]}  } }{Q^{\Omega_1}_{[n]} A^{\Omega_1}_{[n]}}{    \vect{W}^{ \theta_{[\eta],[\lambda]} }  } \\ %
			\geq  \sum_{j=1}^{n}     \MIcond{ \vect{W}^{[f]\setminus\theta_{[\eta],[\lambda]}  } }{Q^{\Omega_1}_{j} A^{\Omega_1}_{j}}{ \vect{W}^{ \theta_{[\eta],[\lambda]} }}
		   \\ 
			\stackrel{(a)}{=} \sum_{j=1}^{n} \MIcond{ \vect{W}^{[f]\setminus\theta_{[\eta],[\lambda]} } }{Q^{\Omega_2}_{j} A^{\Omega_2}_{j}}{ \vect{W}^{ \theta_{[\eta],[\lambda]} }} \\ 
			\stackrel{(b)}{=} \sum_{j=1}^{n}   \MIcond{ \vect{W}^{[f]\setminus \theta_{[\eta],[\lambda]}   } }{ A^{\Omega_2}_{j}}{ Q^{\Omega_2}_{j} \vect{W}^{ \theta_{[\eta],[\lambda]} }  } \\ 
			\stackrel{(c)}{=} \sum_{j=1}^{n} \bigHPcond{ A^{\Omega_2}_{j}}{Q^{\Omega_2}_{j} \vect{W}^{ \theta_{[\eta],[\lambda]} } } 
			-\underbrace{\bigHPcond{ A^{\Omega_2}_{j}}{Q^{\Omega_2}_{j} \vect{W}^{ \theta_{[\eta],[\lambda]} }  \vect{W}^{[f]\setminus \theta_{[\eta],[\lambda]}  } }
			}_{=0}\\
			\geq \sum_{j=1}^{n} \bigHPcond{ A^{\Omega_2}_{j}}{Q^{\Omega_2}_{[n]}A^{\Omega_2}_{[j-1]} \vect{W}^{ \theta_{[\eta],[\lambda]} }  }  \\ 
			\stackrel{(c)}{=}  \sum_{j=1}^{n}  \MIcond{ \vect{W}^{[f]\setminus\theta_{[\eta],[\lambda]}  } }{ A^{\Omega_2}_{j}}{Q^{\Omega_2}_{[n]}A^{\Omega_2}_{[j-1]} \vect{W}^{ \theta_{[\eta],[\lambda]} }  }  
			+ \underbrace{
			 \bigHPcond{ A^{\Omega_2}_{j}}{Q^{\Omega_2}_{[n]}A^{\Omega_2}_{[j-1]} \vect{W}^{ \theta_{[\eta],[\lambda]} }  \vect{W}^{[f]\setminus\theta_{[\eta],[\lambda]}  } }
			}_{=0}\\
			=  \MIcond{\vect{W}^{[f]\setminus\theta_{[\eta],[\lambda]}  }}{ A^{\Omega_2}_{[n]}}{Q^{\Omega_2}_{[n]} \vect{W}^{ \theta_{[\eta],[\lambda]} }  } \\ 
			\stackrel{(b)}{=}  \MIcond{\vect{W}^{[f]\setminus\theta_{[\eta],[\lambda]}  } }{ A^{\Omega_2}_{[n]} Q^{\Omega_2}_{[n]} }{ \vect{W}^{ \theta_{[\eta],[\lambda]} }  } \\
			\stackrel{(d)}{=}   \MIcond{\vect{W}^{[f]\setminus\theta_{[\eta],[\lambda]} }}{ A^{\Omega_2}_{[n]} Q^{\Omega_2}_{[n]} \vect{W}^{ \theta_{[\eta+1:2\eta],[\lambda]}  }}{ \vect{W}^{ \theta_{[\eta],[\lambda]} }  }  
			- \underbrace{
				\MIcond{\vect{W}^{[f]\setminus \theta_{[\eta],[\lambda]} } }{ \vect{W}^{ \theta_{[\eta+1:2\eta],[\lambda]}  }  }{ A^{\Omega_2}_{[n]} Q^{\Omega_2}_{[n]} \vect{W}^{ \theta_{[\eta],[\lambda]} }  }
			 	}_{=0}\\	\\
				=   \MIcond{\vect{W}^{[f]\setminus \theta_{[\eta],[\lambda]}  } }{ \vect{W}^{ \theta_{[\eta+1:2\eta],[\lambda]}  }}{ \vect{W}^{ \theta_{[\eta],[\lambda]} }  }  
				+  \MIcond{\vect{W}^{[f]\setminus \theta_{[\eta],[\lambda]} } }{ A^{\Omega_2}_{[n]} Q^{\Omega_2}_{[n]} }{ \vect{W}^{ \theta_{[\eta],[\lambda]} }  \vect{W}^{ \theta_{[\eta+1:2\eta],[\lambda]}  }  }  \\
				\stackrel{(e)}{=}    \bigHP{\vect{W}^{ \theta_{[\eta+1:2\eta],[\lambda]}  } }   +  \MIcond{   \vect{W}^{[f]\setminus\theta_{[2\eta],[\lambda]} } }{Q^{\Omega_2}_{[n]} A^{\Omega_2}_{[n]}}{\vect{W}^{ \theta_{[2\eta],[\lambda]} }} \nonumber \\
				\stackrel{(f)}{=}  \eta\lambda\const{L}   +  \MIcond{ \vect{W}^{[f]\setminus \theta_{[2\eta],[\lambda]}  } }{Q^{\Omega_2}_{[n]} A^{\Omega_2}_{[n]}}{\vect{W}^{ \theta_{[2\eta],[\lambda]} }}, 
\end{IEEEeqnarray*}
where 
\begin{itemize}
\item[$(a)$] follows from the privacy constraint \eqref{eq:MPPIR-privacy};
\item[$(b)$] follows from the independence between the messages \eqref{eq:JointEntropy} and the  independence between the messages and the queries  \eqref{eq:IndepQM};
\item[$(c)$] follows from the fact that the answer strings are a deterministic function of the queries and the stored messages  \eqref{eq:DeterministicAnswers};
\item[$(d)$] follows from the chain rule of mutual information, the independence of the messages \eqref{eq:JointEntropy}, and the correctness condition  \eqref{eq:MPPIR-correctness}, in particular from $ 	\bigHPcond{ \vect{W}^{ \theta_{[\eta+1:2\eta],[\lambda]}  }}{Q^{\Omega_2}_{[n]} A^{\Omega_2}_{[n]} }=0  $;
\item[$(e)$] follows from the independence of the messages \eqref{eq:JointEntropy};
\item[$(f)$] follows from the chain rule of mutual information and the fact that each message consists of $\const{L}$ independent and identically distributed symbols \eqref{eq:Entropy}.
\end{itemize}
\end{proof}

\section{Proof of Lemma \ref{lem:RemainderLemma}}\label{app:RemainderLemma}
Here, we prove a lower bound on the conditional mutual information stated in  \cref{lem:RemainderLemma}. 
Before starting the proof, recall that

	$$\vect{W}^{ \theta_{[\eta],[\lambda]} } \triangleq \{ \vect{W}^{(\theta_{1,1})}, \vect{W}^{(\theta_{1,2})},\dots, \vect{W}^{(\theta_{1,\lambda})}, \vect{W}^{(\theta_{2,1})},\dots,\vect{W}^{(\theta_{\eta,1})},\dots,\vect{W}^{(\theta_{\eta,\lambda})} \},$$
	and
	$${\vect{W}^{[f]\setminus\theta_{[\eta],[\lambda]}} \triangleq  \vect{W}^{[\theta_{1,\lambda+1}:\theta_{2,1}-1]} \cup \vect{W}^{[\theta_{2,\lambda+1}:\theta_{3,1}-1]} \cup \dots \cup \vect{W}^{[\theta_{\eta-1,\lambda+1}:\theta_{\eta,1}-1]}\cup  \vect{W}^{[\theta_{\eta,\lambda+1}:f]}}.$$ 

	\begin{proof}
	We start the proof with $\Omega_1=[\eta]$ and $\Omega_2=[\Gamma-\eta+1:\Gamma]$. For $\eta\geq \frac{\Gamma}{2}$, we have $\Omega_1\cap\Omega_2=[\Gamma-\eta+1:\eta]$. We then have the following chain of inequalities: 
	\begin{IEEEeqnarray*}{lCr}
		n\MIcond{   \vect{W}^{[f]\setminus \theta_{[\eta],[\lambda]} } }{Q^{\Omega_1}_{[n]} A^{\Omega_1}_{[n]}}{    \vect{W}^{ \theta_{[\eta],[\lambda]} }  } \\ %
		\geq  \sum_{j=1}^{n}     \MIcond{  \vect{W}^{[f]\setminus\theta_{[\eta],[\lambda]}  }  }{Q^{\Omega_1}_{j} A^{\Omega_1}_{j}}{ \vect{W}^{ \theta_{[\eta],[\lambda]} }}
		\\ 
		\stackrel{(a)}{=} \sum_{j=1}^{n} \MIcond{   \vect{W}^{[f]\setminus\theta_{[\eta],[\lambda]} }  }{Q^{\Omega_2}_{j} A^{\Omega_2}_{j}}{ \vect{W}^{ \theta_{[\eta],[\lambda]} }} \\ 
		\stackrel{(b)}{=} \sum_{j=1}^{n}   \MIcond{  \vect{W}^{[f]\setminus \theta_{[\eta],[\lambda]} } }{ A^{\Omega_2}_{j}}{ Q^{\Omega_2}_{j} \vect{W}^{ \theta_{[\eta],[\lambda]} }  } \\ 
		\stackrel{(c)}{=} \sum_{j=1}^{n} \bigHPcond{ A^{\Omega_2}_{j}}{Q^{\Omega_2}_{j} \vect{W}^{ \theta_{[\eta],[\lambda]} } } 
		-\underbrace{\bigHPcond{ A^{\Omega_2}_{j}}{Q^{\Omega_2}_{j} \vect{W}^{ \theta_{[\eta],[\lambda]} }   \vect{W}^{[f]\setminus\theta_{[\eta],[\lambda]} }  }
		}_{=0}\\
		\geq \sum_{j=1}^{n} \bigHPcond{ A^{\Omega_2}_{j}}{Q^{\Omega_2}_{[n]}A^{\Omega_2}_{[j-1]} \vect{W}^{ \theta_{[\eta],[\lambda]} }  }  \\ 
		\stackrel{(c)}{=}  \sum_{j=1}^{n}  \MIcond{ \vect{W}^{[f]\setminus\theta_{[\eta],[\lambda]} } }{ A^{\Omega_2}_{j}}{Q^{\Omega_2}_{[n]}A^{\Omega_2}_{[j-1]} \vect{W}^{ \theta_{[\eta],[\lambda]} }  }  
		+ \underbrace{
			\bigHPcond{ A^{\Omega_2}_{j}}{Q^{\Omega_2}_{[n]}A^{\Omega_2}_{[j-1]} \vect{W}^{ \theta_{[\eta],[\lambda]} }    \vect{W}^{[f]\setminus\theta_{[\eta],[\lambda]} } }
		}_{=0}\\
		=  \MIcond{ \vect{W}^{ [f]\setminus\theta_{[\eta],[\lambda]} } }{ A^{\Omega_2}_{[n]}}{Q^{\Omega_2}_{[n]} \vect{W}^{ \theta_{[\eta],[\lambda]} }  } \\ 
		\stackrel{(b)}{=}  \MIcond{  \vect{W}^{ [f]\setminus\theta_{[\eta],[\lambda]} } }{ A^{\Omega_2}_{[n]} Q^{\Omega_2}_{[n]} }{ \vect{W}^{ \theta_{[\eta],[\lambda]} }  } \\
		\stackrel{(d)}{=}   \MIcond{  \vect{W}^{[f]\setminus \theta_{[\eta],[\lambda]} }   }{ A^{\Omega_2}_{[n]} Q^{\Omega_2}_{[n]} \vect{W}^{ \theta_{[\Gamma-\eta+1:\Gamma],[\lambda]}  }}{ \vect{W}^{ \theta_{[\eta],[\lambda]} }  }  
		- \underbrace{
			\MIcond{ \vect{W}^{[f]\setminus  \theta_{[\eta],[\lambda]} } }{ \vect{W}^{ \theta_{[\Gamma-\eta+1:\Gamma],[\lambda]}  }  }{ A^{\Omega_2}_{[n]} Q^{\Omega_2}_{[n]} \vect{W}^{ \theta_{[\eta],[\lambda]} }  }
		}_{=0}\\	\\
		=   \MIcond{ \vect{W}^{[f]\setminus \theta_{[\eta],[\lambda]} } }{ \vect{W}^{ \theta_{[\Gamma-\eta+1:\Gamma],[\lambda]}   }}{ \vect{W}^{ \theta_{[\eta],[\lambda]} }  }  
		+  \MIcond{  \vect{W}^{ [f]\setminus \theta_{[\eta],[\lambda]}  } }{ A^{\Omega_2}_{[n]} Q^{\Omega_2}_{[n]} }{ \vect{W}^{ \theta_{[\eta],[\lambda]} }  \vect{W}^{ \theta_{[\Gamma-\eta+1:\Gamma],[\lambda]}  }  }  \\
		%
		=   \bigHPcond{\vect{W}^{ \theta_{[\Gamma-\eta+1:\Gamma],[\lambda]}  } }{ \vect{W}^{ \theta_{[\eta],[\lambda]} } }   +  \MIcond{   \vect{W}^{[f]\setminus \theta_{[\Gamma],[\lambda]} } }{Q^{\Omega_2}_{[n]} A^{\Omega_2}_{[n]}}{\vect{W}^{ \theta_{[\Gamma],[\lambda]} }} \nonumber \\
		\stackrel{(f)}{=}    \bigHP{\vect{W}^{ \theta_{[\eta+1:\Gamma],[\lambda]}  } }	+  \underbrace{ \MIcond{   \vect{W}^{[f]\setminus \theta_{[\Gamma],[\lambda]} } }{ A^{\Omega_2}_{[n]}}{ Q^{\Omega_2}_{[n]} \vect{W}^{ \theta_{[\Gamma],[\lambda]} }} 
			}_{=0}\\
		\stackrel{(g)}{=}  \lambda\const{L} (\Gamma-\eta),   
	\end{IEEEeqnarray*}
	where 
	\begin{itemize}
		\item[$(a)$] follows from the privacy constraint \eqref{eq:MPPIR-privacy}; 
		\item[$(b)$] follows from the independence between the messages \eqref{eq:JointEntropy} and the  independence between the messages and the queries  \eqref{eq:IndepQM};
		\item[$(c)$] follows from the fact that the answer is a deterministic function of the queries and the stored messages  \eqref{eq:DeterministicAnswers};
		\item[$(d)$] follows from the chain rule of mutual information, the independence of the messages \eqref{eq:JointEntropy}, and the correctness condition  \eqref{eq:MPPIR-correctness}, in  particular from $ 	\bigHPcond{ \vect{W}^{ \theta_{[\Gamma-\eta+1:\Gamma],[\lambda]}  }}{Q^{\Omega_2}_{[n]} A^{\Omega_2}_{[n]} }=0  $;
		
		\item[$(f)$] follows from the independence of the messages \eqref{eq:JointEntropy}; the second term is zero due to the independence of the messages and the queries \eqref{eq:IndepQM} and the fact that the answer strings  are a deterministic function of the queries and a \emph{sufficient} number of messages from each classes. Specifically, by combining 
		\eqref{eq:DeterministicAnswers} and  \eqref{eq:AnswersFromPartialDS} we have 
		\begin{IEEEeqnarray*}{lCr}
			\bigHPcond{A^{\Omega_2}_{[n]}}{Q^{\Omega_2}_{[n]} \vect{W}^{ \theta_{[\Gamma],[\lambda]}  } }\\
			=	\bigHPcond{A^{\Omega_2}_{[n]}}{Q^{\Omega_2}_{[n]} \vect{W}^{[\theta_{1,1}:\theta_{1,\lambda}]}\vect{W}^{ \theta_{[2:\Gamma],[\lambda]} } } \\
			= \bigHPcond{A^{\Omega_2}_{[n]}}{Q^{\Omega_2}_{[n]} \vect{W}^{[\theta_{1,1}:\theta_{2,1}-1]} 	\vect{W}^{[\theta_{2,1}:\theta_{2,\lambda}]}  \vect{W}^{ \theta_{[3:\Gamma],[\lambda]}} } \\
			=	\vdots\\
			= \bigHPcond{A^{\Omega_2}_{[n]}}{Q^{\Omega_2}_{[n]}\vect{W}^{[f]}}=0;
		\end{IEEEeqnarray*}  
		
		\item[$(g)$] follows from the fact that each message consists of $\const{L}$ independent and identically distributed symbols \eqref{eq:Entropy}.
	\end{itemize}
\end{proof}

%



\bibliographystyle{IEEEtran} 
\bibliography{defshort.bib,biblioHY.bib,references.bib}

\end{document}